\pdfoutput=1
\documentclass[12pt,draftclsnofoot, onecolumn]{IEEEtran} 
\usepackage{amssymb}
\usepackage{amsthm}
\usepackage{graphicx}
\usepackage{array,color}
\usepackage{amsmath}
\usepackage{stfloats}
\usepackage{graphicx}
\usepackage{subfigure}
\usepackage{tabularx}
\usepackage{epsfig,epsf,color,balance,cite}
\usepackage{algorithmic}
\usepackage{algorithm}
\usepackage{bm}
\usepackage{textcomp}
\usepackage{color}
\usepackage{multirow}
\usepackage{cite}
\usepackage{enumerate}
\usepackage{cases}
\usepackage{color}
\usepackage{url}
\usepackage{epstopdf}

\begin{document}
\title{Cost Minimization for Cooperative Computation Framework in MEC Networks}
\author{\IEEEauthorblockN{Yijin Pan, Cunhua Pan, Kezhi Wang, Huiling Zhu, and Jiangzhou Wang,~\IEEEmembership{Fellow,~IEEE}}
\thanks{Y.Pan are with the National Mobile Communications Research Laboratory, Southeast University, Nanjing 211111, China. She is also with School of Engineering and Digital Arts, University of Kent, UK. Email: panyj@seu.edu.cn, y.pan@kent.ac.uk.}
\thanks{C.Pan is with the School of Electronic Engineering and Computer Science, Queen Mary, University of London, London E1 4NS, UK. Email: c.pan@qmul.ac.uk.}
\thanks{K.Wang is with the Department of Computer and Information Sciences, Northumbria University, UK. Email: kezhi.wang@northumbria.ac.uk.}
\thanks{J.Wang and H.Zhu are with the School of Engineering and Digital Arts, University of Kent, UK. Email: J.Z.Wang@kent.ac.uk, H.Zhu@kent.ac.uk.}}

\maketitle

\vspace{-2em}
\begin{abstract}
In this paper, a cooperative task computation framework exploits the computation resource in UEs to accomplish more tasks meanwhile minimizes the power consumption of UEs.
The system cost includes the cost of UEs' power consumption and the penalty of unaccomplished tasks,  and the
system cost is minimized by jointly optimizing binary offloading decisions, the computational frequencies, and the offloading transmit power.
To solve the formulated mixed-integer non-linear programming problem, three efficient algorithms are proposed, i.e., integer constraints relaxation-based iterative algorithm (ICRBI), heuristic matching algorithm, and the decentralized algorithm.
The ICRBI algorithm achieves the best performance at the cost of the highest complexity, while the heuristic matching algorithm significantly reduces the complexity while still providing reasonable performance.
As the previous two algorithms are centralized, the decentralized algorithm is also provided to further reduce the complexity, and it is suitable for the scenarios that cannot provide the central controller.
The simulation results are provided to validate the performance gain in terms of the total system cost obtained by the proposed cooperative computation framework.
\end{abstract}

\newpage

\vspace{-1em}
\section{Introduction}


The 5G mobile communication network and its future evolution (5G Beyond, B5G) are going to brace various unprecedented services such as automatic driving and the Internet of Things (IoT), which demand high computation resources and ultra-low-latency communications. 
To support these computation-intensive and delay-sensitive applications, mobile edge computing (MEC) is introduced as a key technology of the B5G communication system\cite{mao2017survey}.
In the MEC network, the computing server is deployed at the network edge, enabling users to offload their computing tasks to the MEC server for execution.

\vspace{-1em}
\subsection{User Cooperative Computing}

According to the forecast from the industry, there will be 12.3 billion mobile-connected devices by 2022 \cite{index2019global}.
It can be foreseen that the relatively limited computation capability of the MEC server has to be shared by the intensive workloads from the massive IoT devices\cite{8166725}.
However, due to the extensive computation-intensive applications, it may be not sufficient to provide satisfactory computation services only relying on the computation resource of the MEC servers.
In light of the rapid development of these smart devices, the cooperative computation is proposed as a promising way to enhance the MEC service\cite{tran2017collaborative}.
In the cooperative computation, the computation resources within UEs can be shared via device-to-device (D2D) communications \cite{li2014exploring}.
By introducing the cooperative computation, a large amount of computing resource can be harvested without additional costs of network access facilities\cite{tang2018enabling}.
To efficiently exploit the distributed computation resources within UEs, a number of efforts have been devoted to the design of cooperative computing resource management, e.g. \cite{Wang2018, FDMA2017AL}.

{Most of the current contributions on the cooperative computation have been dedicated to improving the power efficiency and the delay performance.
From the perspective of power efficiency, a number of studies targeted at maximizing the power efficiency or minimizing the energy consumption\cite{sheng2018energy,huang2019bilevel}, as the cooperative MEC networks are usually constrained by the limited power of UEs\cite{altman2019forever}.}
By offloading tasks to nearby UEs for execution, the power required for task offloading can be significantly reduced\cite{chen2017exploiting,song2014energy}.
In \cite{cao2018joint}, UE can act as a computation helper or a relay to reduce UE's energy consumption.
In \cite{8403705}, the helper's CPU idling time was predicted so that the available computing resources can be fully utilized.
{From the perspective of reducing delay, }multiple UEs were employed in \cite{xing2019joint} by the time division multiple access (TDMA) protocol to help reduce the delay.
The average delay minimization problem for the multi-hop ad hoc networks was investigated in \cite{MengyuanWang.2017}.
Furthermore, in \cite{Diao.2019}, the energy consumption and delay of all users in the cooperative MEC system was jointly minimized.

{Another important benefit is improving the computation capacity by leveraging the cooperative computation.}
Due to the limited computation capability of the MEC server, some tasks may not be accomplished, resulting in the so-called infeasible tasks. 
In cooperative computation, the transmission delay for task offloading can be significantly reduced \cite{wang2017delay}.
As a result, the previous infeasible tasks (cannot be accomplished in time) can become feasible by exploiting the D2D based cooperative computation.

\vspace{-0.75em}
\subsection{Motivations}

{In practice, a very challenging scenario often occurs in the cooperative computation is that UEs have their own tasks to be executed meanwhile it still has the potential to help others.
However, this issue has been ignored by a lot of current researches. 
For instance, specific idle UEs dedicated to helping computation was assumed in \cite{cao2018joint,xing2019joint, Qiao.2019,sheng2018energy} for simplicity.
The cooperative UE pairs were pre-determined in \cite{lin2019optimal,You.2018,Gu.2018} to avoid the complicated task scheduling.  
Although a similar scenario has been considered in \cite{he2019d2d,kai2019energy,liu2018d2d}, some key features  still have not been well addressed.
In fact, a task executed by cooperative computation is accomplished at the expense of power costs from both the task owner and the computation helper.
Moreover, the allocated computing frequency to the offloaded task also affects the power required for offloading transmission.
Meanwhile, UEs are power-limited and computation resources-limited.
However, these power consumption constraints of UEs were ignored in the above-mentioned literatures\cite{liu2018d2d,kai2019energy}, which greatly restricts the practicability of the proposed approaches.
The approach in \cite{he2019d2d} targeted to maximize the number of accomplished tasks, but for the tasks that are highly integrated or relatively simple, the proposed partial offloading scheme in \cite{he2019d2d} is not applicable.}

{Accomplishing more tasks in the cooperative computation requires higher power consumption from UEs.
Consequently, without a proper coordination of UEs' cooperation, the power efficiency can be degraded greatly, and the number of accomplished tasks can be reduced.
Thus, exploiting the trade-off between the achieved computation capacity and the power consumption of UEs is crucial, however, which unfortunately has not been well addressed in the existing literature.
Therefore, we aim to reveal the cooperative computation solution that can complete the most tasks while consuming the least power, which is different from the existing work.}

{Considering the fact that power budgets and computation capabilities are different by UEs, we introduce the system cost to better understand the trade-off between the number of accomplished tasks and the power consumption.
To be specific, in our work, the system cost is modelled as the cost of UEs' power consumption and the penalty for unaccomplished tasks.
Different from all the above works, we consider a general cooperative computation framework, where each UE has its own task to be executed with its delay requirement. 
The tasks of UEs cannot be divided and can be offloaded to MEC server or nearby UEs for execution.
The binary offloading decisions, i.e., which task should be offloaded to which UE or MEC, is optimized together with the power control and computation frequency assignment, which are determined by the formulated intractable mixed integer non-linear problem (MINLP).
As the binary offloading decision of one UE affects the delay and power consumption of all other UEs}, solving the formulated MINLP problem is quite challenging.

\vspace{-1em}
\subsection{Contributions and Organization}

The main contributions of this paper are summarized as follows:
\begin{itemize}
\item {We first present a proposition to transform the original intractable MINLP problem into an equivalent tractable form. 
Then, we provide another proposition to reveal the feasibility set of devices for a given task}, so that the feasible region for obtaining the optimal solution can be efficiently reduced. 

\item  {Based on the propositions, an integer constraints relaxation-based iterative (ICRBI) algorithm is first proposed to efficiently solve the MINLP problem by leveraging the Lagrange dual method and the Karush-Kuhn-Tucker (KKT) conditions. Regarding the high computation complexity of ICRBI, the low-complexity heuristic matching algorithm is proposed to speed up the convergence, and task matching criteria targeting at maximizing the number of accomplished tasks and minimizing the power costs are provided to guarantee reasonable performance.}
{In addition, as the ICRBI algorithm and heuristic matching algorithm are both centralized, a decentralized algorithm is also provided to avoid centralized coordination and further reduce the complexity.}

\item Finally, the simulation results are presented to validate the performance gain achieved by the proposed cooperative computation schemes.

\end{itemize}

Although in our previous conference version \cite{panpower}, we have shown that the cooperative computation scheme allows more tasks to be accomplished, the comprehensive problem analysis, the low-complexity heuristic algorithms, and the decentralized algorithm are not provided in \cite{panpower}.

\section{System Model and Problem Formulation}

\subsection{System Model}

Consider that a network access point (AP) equipped with an MEC server is serving $N$ UEs for computation service. 
{In our MEC system, we assume that the UEs associated with the same AP can help each other for task computation via D2D transmissions.
Let $\mathcal{M}=\{0,1,2,\cdots,N\}$ denote the set of devices that UEs can offload tasks to, which includes the MEC server (index $0$) and all the $N$ UEs.
In the following, we adopt the term `` device '' to refer to the elements in set $\mathcal{M}$.}
{Assume that the devices adopt Frequency Division Duplexing (FDD) to enable transmitting and receiving tasks at the same time, and D2D communications work in the overlaying mode.}
\footnote{{Through proper channel allocation, such as the graph-colouring based channel assignment, a certain degree of channel multiplexing gain in D2D communications can be obtained and the interference can be limited to the same level as the background noise so that D2D interference caused by channel multiplexing is included in the noise.} }

Assume that each UE has a computation task to be executed, and the task of UE $i$ is called ``task $i$'' in the following descriptions.
The task set is denoted as $\mathcal{N}=\{1, 2, \cdots, N\}$.
{The requirements of computation tasks are different by UEs.}
Then, the computational intensive task $i$ can be modeled as $(F_i, D_i, T_{i}^{max}), i\in \mathcal{N}$, where $F_i$ {(CPU cycles)} is the required CPU cycles of task $i$ for computation, $D_i$ {(bits)} denotes the data size of task $i$ for transmitting and $T_{i}^{max}$ {(seconds)} is the latency constraint of task $i$.

\begin{figure}
	\centering
	\vspace{-2em}
	\includegraphics[width=0.55\textwidth]{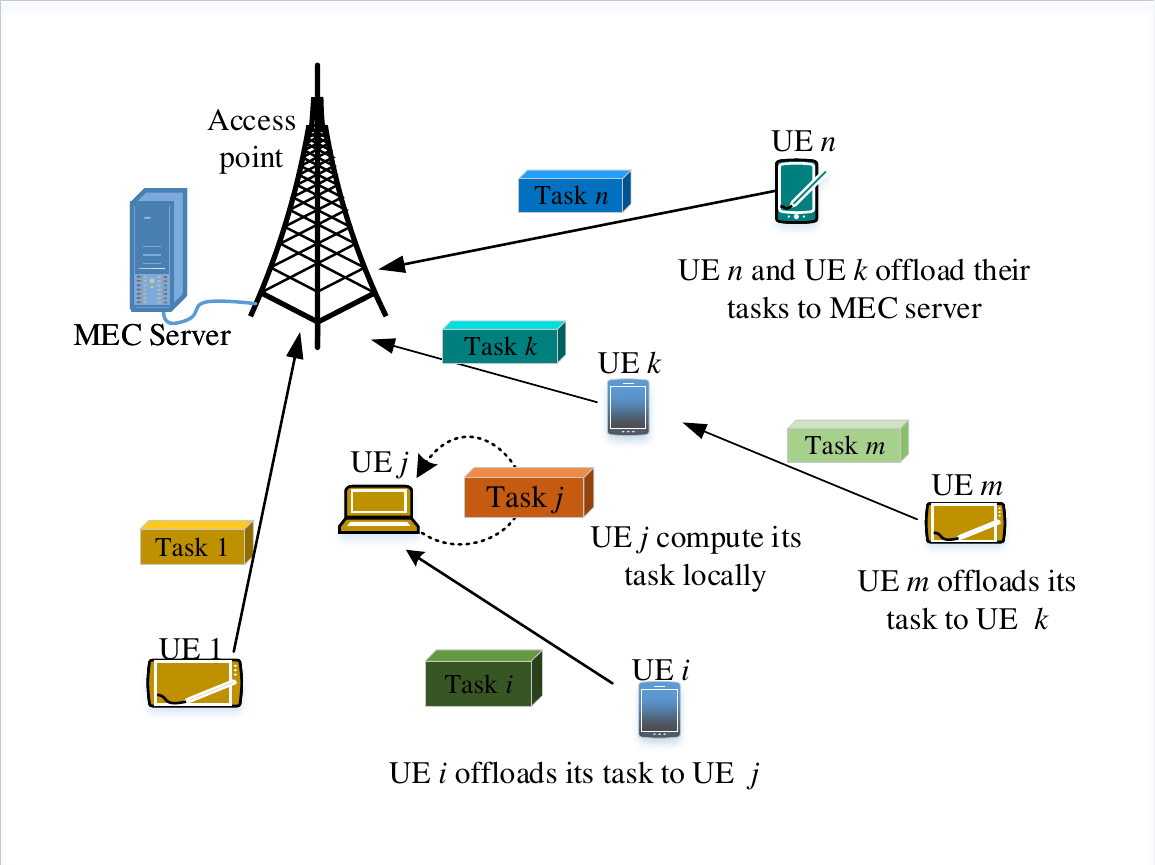}
	\vspace{-2em}
	\caption{{A possible realization of the cooperative computation framework.}} \label{fig0}
	\vspace{-2em}
\end{figure}

Fig. \ref{fig0} illustrates a possible realization of the cooperative task computation.
{We only consider the D2D transmission with ``one hop", i.e., the device received the offloaded task will not continue to offload it.}
In Fig. \ref{fig0}, the computation capacity of the MEC server may be limited so that all the $N$ tasks may not be successfully accomplished simultaneously.
In Fig. \ref{fig0}, UE $i$ can offload its task to UE $j$, which has a higher computing capability. 
{While helping UE $i$, UE $j$'s own task should be executed in time.
When the computation resources of UE $j$  is sufficient, as shown in Fig. \ref{fig0}, task $j$ can be executed locally.  
Otherwise, for instance, the computation resources of UE $k$ is not sufficient to cope with both task $k$ and task $m$ at the same time.
However, the power budget of UE $k$  is sufficient to offload its own task to the MEC server, and compute the offloaded tasks from UE $m$. 
The offloading decision of each task affects the offloading decisions of all other tasks.
As the UEs are resource-limited and the tasks are delay-constrained, finding efficient and feasible offloading decisions for all tasks becomes very challenging. }

Let us use the indicator $a_{i,j}$, $i\in \mathcal{N}$, $j\in \mathcal{M}$ to represent the decisions of task offloading, and
\begin{equation}\label{w1}
C1: a_{i,j} =\{ 0,1\}, \forall i\in  \mathcal{N}, \forall j\in  \mathcal{M}.
\end{equation}
That is to say, $a_{i,j}=1$ denotes that task $i$ is offloaded to UE $j$ ($j \neq 0$), or to the MEC server ($j=0$). 
In addition, we assume that each task can only be executed in one device:
\begin{equation}\label{w2}
C2: \sum_{j\in \mathcal{M}}a_{i,j} \leq 1, i\in \mathcal{N}.
\end{equation}
Note that some tasks may not be able to be accomplished anywhere in the required time due to the lack of communication or computation resources.
{
To effectively avoid unnecessary communication costs,  the offloading operation will be conducted after the offloading decisions are made.
As the offloading decisions will be optimized, only the tasks that can be accomplished successfully will be assigned an execution device, i.e., there exists a $a_{i,j} =1, \forall j \in \mathcal{M}$, then task $i$ is regarded as an ``accomplished task''. }

If UE $i$ offloads its task $i$ to device $j$, the achievable transmission data rate {(bps)} is
\begin{equation}\label{w4}
\begin{aligned}
r_{i,j}= B \log_2\left(1+\frac{p_{i,j}^{T} h_{i,j}}{\sigma^2}\right),  \forall i\in \mathcal{N},  j\in \mathcal{M},
\end{aligned}
\end{equation}
where all UEs are allocated with orthogonal frequency bands and equal bandwidths, $h_{i,j} $ represents the channel gain {from} UE $i$ to device $j$, $\sigma^2$ {(watt)} describes the white Gaussian noise power, $B$ {(Hz)} denotes the allocated bandwidth and $p_{i,j}^{T}$ {(watt)} is the transmit power.
Then, the time for task offloading transmission is $T_{i,j}^{T}=\frac{D_i}{r_{i,j}}, \forall i\in \mathcal{N}, j\in \mathcal{M}, i\neq j.$

In addition, the execution time of task $i$ is $T_{i,j}^C=\frac{F_i}{f_{i,j}}, \forall i\in \mathcal{N}, \forall j\in \mathcal{M}$, where $f_{i,j}$ is the computation speed (in CPU cycles per second) provided by device $j$ to execute task $i$. 
The total time consumption should satisfy the latency constraint:
\begin{equation}\label{w8}
\begin{aligned}
C3: \sum_{j \neq i, j \in \mathcal{M}} a_{i,j} \left(T_{i,j}^{T} +T_{i,j}^C \right ) + a_{i,i}T_{i,i}^C\leq T_{i}^{max}, i\in \mathcal{N}.
\end{aligned}
\end{equation}

In addition, the computational frequency of each device should not exceed its maximum computational capacity: 
\begin{equation}\label{www8}
\begin{aligned}
C4: \sum_{i \in \mathcal{N}}a_{i,j} f_{i,j} \leq f_j^{max}, j\in \mathcal{M},
\end{aligned}
\end{equation}
where $f_j^{max}$ is the maximum CPU capacity of device $j$.
{With the computational capacity constraints $C4$ and the delay constraints $C3$, the number of tasks that each UE or the MEC server can support for successful execution is limited. }

{According to \cite{Burd.1996}}, the computing power consumption for device $j$ to execute task $i$ with computational speed $f_{i,j}$ can be modelled as
\begin{equation}\label{ww8}
\begin{aligned}
p_{i,j}^{C}= \kappa_j(a_{i,j}f_{i,j})^{\nu_j},  \forall j\in \mathcal{N},
\end{aligned}
\end{equation}
where $\kappa_j \geq 0$ {(Joule/$(\text{CPU operations})^{\nu_j-1}$)} is the effective switched capacitance and $\nu_j \geq 1$ is a positive constant.
The value of $\nu_j$ depends on the CPU chip structure.

{Furthermore, mobile UEs are power-limited.
According to the power consumption models in \cite{Zhou.2014,Zhou.2017,Wang.20069282006928,Jiang.2016,Wu.2017}, 
the power consumption of UE $i$'s transmitter and receiver consists of two parts: the first is the power consumed in the power amplifier (PA) for transmission, which is proportional to the transmit power $p_{i,j}^T$; the second is the static power consumed for circuits including the mixer, filters, frequency synthesizer, D/A and D/A converter, etc., which can be modelled as a constant.
Then, the power consumption of UE $i$'s transmitter and receiver can be expressed as
\begin{equation}
p_i^{TR}  = \frac{1}{\eta_i} \sum_{j \neq i, j \in \mathcal{M}} a_{i,j} p_{i,j}^T + p_i^{cir}
\end{equation}
where $p_i^{cir}$ is the total circuit power consumption of receiver and transmitter, and $\eta_i$ is the PA efficiency, i.e., $0 <\eta_i < 1$.}
Then, one has the power constraint for UE $i$ as
\begin{equation}\label{c5}
\begin{aligned}
{C5: p_i = \sum_{k \in \mathcal{N}} a_{k,i} p_{k,i}^C +  \frac{1}{\eta_i} \sum_{j \neq i, j \in \mathcal{M}} a_{i,j} p_{i,j}^T + p_i^{cir} \leq p_i^{max}, i\in \mathcal{N}},
\end{aligned}
\end{equation}
where { $p_i^{max}$ is the maximum power budget of UE $i$.}
{In the maximum power constraint $C5$, the total power consumption includes that of the computation, UE's transmitter, and receiver.}
Note that the network access point is normally cable powered, so that the power constraint for the MEC server can be neglected.

\vspace{-1em}
\subsection{Problem Formulation}

Note that the mobile UEs are generally power limited, and their power budgets and hardware costs are various due to the heterogeneity of hardware capabilities.
Based on this concern, we introduce a price parameter to charge the power costs of UEs.
Define the price for the unit power of UE $i$ as $w_i$ {(price per watt)}.
{ According to (\ref{c5}), the cost of UE $i$'s total power consumption denoted by $\Psi_{i}$ is given by} 
\vspace{-0.75em}
\begin{equation}\label{ci}
{\Psi_{i} = w_i\left(\sum_{k \in \mathcal{N}} a_{k,i} p_{k,i}^C +\frac{1}{\eta_i} \sum_{j \neq i, j \in \mathcal{M}} a_{i,j} p_{i,j}^T + p_i^{cir}\right)}.
\end{equation}

When the number of unaccomplished tasks increases, the quality of user experience will be degraded, and many complaints will be received at the network operator.
{As a result, we define a positive constant $\phi_i$ {(price per task)} to capture the penalty for the unaccomplished task $i$, and its value depends on the practical requirements of the network operators and the task itself.
The penalty of task $i$ is denoted by $\Delta_i$, which can be expressed as}
	\begin{equation}
	{\Delta_i = \phi_i\left(1-\sum_{j\in \mathcal{M}}a_{i,j}\right). \label{Penalty}}
	\end{equation}
{Introducing the penalty can push the users to participate in cooperative computing.
In addition, adjusting penalties for different tasks can help the network adjust priorities of tasks.
For instance, if the user $i$'s task was left uncompleted in the previous slot, its penalty $\phi_i$ should increase in the current slot for compensation and fairness.}

Then, the objective $\mathcal{C}$ is defined as the total system cost, which consists of the power cost capturing the power consumption of UEs and the penalty for the unaccomplished tasks, i.e., $\mathcal{C}_{total}= \sum_{i\in \mathcal{N}}\Psi_i + \sum_{i\in \mathcal{N}} \Delta_i$. According to (\ref{ci}) and (\ref{Penalty}), system cost can be formulated as:
\begin{equation}
\mathcal{C}_{total}= {\sum_{i\in \mathcal{N}} \sum_{j \neq i, j \in \mathcal{M}} \frac{w_i}{\eta_i} a_{i,j} p_{i,j}^T 
	+ \sum_{i\in \mathcal{N}}\sum_{k \in \mathcal{N}} w_i a_{k,i} p_{k,i}^C  + \sum_{i\in \mathcal{N}}w_ip_i^{cir}
	+ \sum_{i\in \mathcal{N}}\phi_i\left(1 - \sum_{j\in \mathcal{M}}a_{i,j}\right)}. \label{cc}
\end{equation}

By optimizing the task offloading decision $\{a_{i,j}\}$, transmit power for task offloading $\{p_{i,j}^T\}$ and the serving computation speed $\{f_{i,j}\}$, the total system cost $\mathcal{C}_{total}$ can be minimized, so that the most tasks can be accomplished with the least cost of power consumption.
Then, the problem can be formulated as
\vspace{-1em}
\begin{subequations}\label{pro1}
\begin{align}
\underset{\{a_{i,j}\},\{f_{i,j}\}, \{p_{i,j}^T\}}{\text{min}}  
&\quad \mathcal{C}_{total} \label{obj1}  \\
\text{s.t.} & \quad C1 - C5. 
\end{align}
\end{subequations}

It is ready  to see that Problem (\ref{pro1}) is an MINLP, which is non-convex and hard to solve in general.
{
In the following section, we first obtain two propositions to help solve this problem.
In the first Proposition, the nonconvex constraints are transformed into tractable forms, so that an equivalent but tractable form of the problem is obtained.
Then, in Proposition \ref{pro_2}, for a given task, the set of devices that cannot successfully execute this task is identified, so that the search region for finding the optimal solution can be sufficiently reduced.  }

\section{Problem Analysis}

\newtheorem{proposition}{\textbf{Proposition}}
\newtheorem{definition}{\textbf{Definition}}

{According to (\ref{cc}), as $\sum_{i\in \mathcal{N}}w_ip_i^{cir} + \sum_{i\in \mathcal{N}}\phi_i$ are constant so that they can be removed from the objective (\ref{obj1}) without affecting the optimal solution.}
Then, an equivalent reformulation of Problem (\ref{pro1}) is given by
\begin{subequations}\label{pro2}
	\begin{align}
\underset{ \underset{\{f_{i,j}\}}{\{a_{i,j}\},}}{\text{min}} \quad  
	& {\mathcal{C}=\sum_{i\in \mathcal{N}}\sum_{ \underset{j \in \mathcal{M}}{j \neq i}}\frac{w_i}{\eta_i} a_{i,j} U_{i,j}(f_{i,j})  
	+\sum_{i\in \mathcal{N}} \sum_{k \in \mathcal{N}}w_i\kappa_i (a_{k,i}f_{k,i})^{\nu_i} - \sum_{i\in \mathcal{N}} \sum_{j\in \mathcal{M}}\phi_i a_{i,j}} \label{obj2}  \\
	\text{s.t.} 
	&  {\sum_{j \neq i, j\in \mathcal{M}}  \frac{a_{i,j}}{\eta_i}  U_{i,j}(f_{i,j}) + \kappa_i \sum_{k \in \mathcal{N}} (a_{k,i} f_{k,i})^{\nu_i} \leq  p_i^{m} , i  \in \mathcal{N}, \label{Pimax}} \\	
	&   a_{i,j}f_{i,j} \geq a_{i,j}f_i^{min}, i  \in \mathcal{N}, j  \in \mathcal{M}, \label{afmax}\\
	&    C1,C2,C4,  \nonumber
	\vspace{-1em}
	\end{align}
\end{subequations}
where $p_i^{m} = p_i^{max}- p_i^{cir}$, $f_i^{min} = \frac{F_i}{T^{max}_{i}}$.
{The transmit power $p_{i,j}^T$ is represented as a function of computation speed $f_{i,j}$ as $p_{i,j}^T = U(f_{i,j})$, and 
$U_{i,j}(x) = \frac{\sigma^2}{h_{i,j}} \left(\exp\left( \frac{\ln 2}{B} \frac{D_i x}{T_i^{max}x-F_i}\right)-1\right)$.
 First, Proposition \ref{pro_1} is provided to show that Problem (\ref{pro2}) and Problem (\ref{pro1}) are equivalent.}

\begin{proposition} \label{pro_1}
Problem (\ref{pro2}) and Problem (\ref{pro1}) are equivalent in the sense that the global optimal objective values of the two problems are identical. 
\end{proposition}
\noindent\textbf{\textit{Proof}}:
{Refer to Appendix \ref{App1} for the detailed proof.}
\qed

{Then, to efficiently solve Problem (\ref{pro2}), Proposition (\ref{pro_2}) is introduced to identify the set of users that cannot successfully execute a specific task, i.e., the infeasible device set of task $i$.}
\begin{proposition} \label{pro_2}
All the device $j$ in the set $\mathcal{J}_{i}$ cannot successfully execute task $i$, where the set $\mathcal{J}_{i}$ is given by
\vspace{-1em}
\begin{equation} \label{Jij}
\mathcal{J}_{i} = \left\{ j \left| f_{i,j}^{D} \geq f_{i,j}^{U} \text{ or } T_i^{max} \leq \frac{D_i}{R_{i,j}^{max} }, j \in \mathcal{M} \right.\right \},
\end{equation}
where $f_{i,j}^{U}$, $f_{i,j}^{D}$ and $R_{i,j}^{max}$ are given in (\ref{Up}) and (\ref{Dn}), respectively.
\end{proposition}
\noindent\textbf{\textit{Proof}}:
{Refer to Appendix \ref{App2} for the detailed proof.}
\qed

\section{Integer Constraints Relaxation Based Algorithm}

In this section, based on Proposition \ref{pro_1} and Proposition \ref{pro_2}, we develop an iterative algorithm by leveraging the relaxation of integer constraints and the Lagrange dual method to efficiently solve the equivalent Problem (\ref{pro2}).  
By introducing the variable $x_{i,j} = a_{i,j} f_{i,j}$, and temporarily relax the integer constraints, Problem (\ref{pro2}) is transformed to 
\begin{subequations}\label{pro3}
	\begin{align}
\underset{\underset{\{a_{i,j}\}}{\{x_{i,j}\}, }}{\text{min}} \quad
	& \mathcal{C}={\sum_{i\in \mathcal{N}}\sum_{ \underset{j \in \mathcal{M}}{j \neq i}}\frac{w_i}{\eta_i} a_{i,j} U_{i,j}\left(\frac{x_{i,j}}{a_{i,j}} \right)  
	+\sum_{i\in \mathcal{N}} \sum_{k \in \mathcal{N}}w_i\kappa_i (x_{k,i})^{\nu_i} - \sum_{i\in \mathcal{N}} \sum_{j\in \mathcal{M}}\phi_i a_{i,j} }\label{obj3} \\
	\text{s.t.} 
	& {\sum_{j \neq i, j\in \mathcal{M}}\frac{a_{i,j}}{\eta_i} U_{i,j}\left(\frac{x_{i,j}}{a_{i,j}} \right)+ \kappa_i \sum_{k \in \mathcal{N}}\left( x_{k,i}\right)^{\nu_i} \leq p^{m}_i , i  \in \mathcal{N}, }  \label{st1}  \\
	&   \sum_{i \in \mathcal{N}}x_{i,j} \leq f_j^{max}, j\in \mathcal{M},   \label{st2}  \\
	&   \sum_{j\in \mathcal{M}}a_{i,j} \leq 1, i\in \mathcal{N},  \label{st3} \\
	&  a_{i,j}f_{i,j}^{D} \leq x_{i,j} \leq a_{i,j}f_{i,j}^{U},  0 \leq a_{i,j} \leq 1, i\in \mathcal{N}, j\in \mathcal{M}.
	\end{align}
\end{subequations}

	
According to (\ref{gfun})-(\ref{PT}), $H_{i,j}(x)$ is a nondecreasing convex function with respect to (w.r.t) $x$, and $G_i(x)$ is convex.
As a result, $U_{i,j}\left( x\right)$ is convex w.r.t $x$, and its perspective function $tU_{i,j}\left( x/t \right)$ is convex w.r.t $(x,t)$.
Then, it is concluded that Problem (\ref{pro3}) is convex, which can be optimally solved by the dual method.
The Lagrangian function of Problem (\ref{pro3}) is given by
\vspace{-0.5em}
\begin{multline}
{\mathcal{L} = \sum_{i\in \mathcal{N}} \sum_{j \neq i, j \in \mathcal{M}}  \!\!\frac{a_{i,j}(w_i+\mu_i)}{\eta_i}U_{i,j}\left( \frac{x_{i,j}}{a_{i,j}}\right)+\sum_{i\in \mathcal{N}}\sum_{k \in \mathcal{N}} (w_i+\mu_i ) \kappa_i \left( x_{k,i}\right)^{\nu_i}
- \sum_{i\in \mathcal{N}} \sum_{j\in \mathcal{M}}\phi_i a_{i,j} } \\
{+ \sum_{j \in \mathcal{M}} v_j \left(\sum_{i \in \mathcal{N}}x_{i,j} - f_{j}^{max}\right) 
- \sum_{i \in \mathcal{N}}\mu_i p^{m}_i 
+ \sum_{i \in \mathcal{N}} s_ i \left(\sum_{j\in \mathcal{M}}a_{i,j}-1\right)},
\end{multline}
where $\mu_i$,$v_ j$ and $s_ i$ are the non-negative dual variables associated with constraints (\ref{st1}),  (\ref{st2}), and (\ref{st3}), respectively.
Then, taking the first-order derivatives of $\mathcal{L}$ w.r.t $x_{i,j}$ and $a_{i,j}$ respectively:
\begin{align}
{\frac{\partial \mathcal{L} }{\partial x_{i,0}}} =&{\frac{(w_i+ \mu_i)}{\eta_i} U'_{i,0}\left(\frac{x_{i,0}}{a_{i,0}}\right) +v_0,  i\in \mathcal{N},} \label{Lxij0}\\
{\frac{\partial \mathcal{L} }{\partial x_{i,j}} }
=& {\frac{(w_i+ \mu_i)}{\eta_i} U'_{i,j}\left(\frac{x_{i,j}}{a_{i,j}}\right) +v_j + (w_j+ \mu_j)\kappa_j \nu_i \left(x_{i,j} \right)^{\nu_j -1}, \forall i \neq j, i,j\in \mathcal{N}, \label{Lxij}} \\
{\frac{\partial \mathcal{L} }{\partial x_{i,i}} }
=&  {(w_i+ \mu_i)\kappa_i \nu_i \left(x_{i,i} \right)^{\nu_i -1} +v_i ,  i\in \mathcal{N},   \label{Lxii}}\\
{\frac{\partial \mathcal{L} }{\partial a_{i,j}}} 
=& {\frac{(w_i+ \mu_i)}{\eta_i} \left( U_{i,j}\left(\frac{x_{i,j}}{a_{i,j}}\right) - \frac{x_{i,j}}{a_{i,j}} U_{i,j}'\left(\frac{x_{i,j}}{a_{i,j}}\right)\right)
- \phi_i +s_i, \forall i \neq  j , i, \in \mathcal{N}, j \in \mathcal{M},  \label{Laij}}\\
{\frac{\partial \mathcal{L} }{\partial a_{i,i}}} =& {- \phi_i + s_i , i \in \mathcal{N},} \label{Laii}
\end{align}
where $U'_{i,j} (x)$ represents the first-order derivative of $U_{i,j}(x)$ w.r.t $x$, given by 
\begin{equation}
U'_{i,j} (x)= \frac{\sigma^2\ln 2}{B h_{i,j}}\exp\left(\frac{\ln 2 D_i x}{B(T_i^{max}x-F_i)} \right)   \frac{-D_i F_i}{(T_i^{max}x-F_i)^2}. \label{ufdiv}
\end{equation}
Next, by applying (\ref{gfun})-(\ref{PT}), the second-order derivative of $U_{i,j}(x)$ w.r.t $x$ is given by 
\begin{equation}
U''_{i,j} (x) = \frac{-U'_{i,j} (x)}{(T_i^{max}x-F_i)} \left(\frac{\ln 2 D_i F_i}{B(T_i^{max}x-F_i)} +T_i^{max}\right) . \label{Udiv2}  
\end{equation}
For $x \in \left[f^{min}_i, + \infty \right)$, it is readily to verify that $U'_{i,j} (x)<0$, $U''_{i,j} (x) > 0 $.
{
In addition,  $\nu_j$ is chosen to be $3$ according to \cite{Burd.1996}.
Then, by taking the second order derivatives of $ \mathcal{L}$ w.r.t $x_{i,j}$ and w.r.t $x_{i,0}$, respectively, we have
$\frac{\partial \mathcal{L}^2 }{\partial^2 x_{i,0}} > 0$ and $\frac{\partial \mathcal{L}^2 }{\partial^2 x_{i,j}}>0$.}
In addition, we have
\begin{align}
{\lim_{x \to {f^{min}_i}^+ } U'_{i,j} (x)+ \frac{\kappa_j \nu_i \eta_i(w_j+ \mu_j)}{w_i+ \mu_i} \left(x \right)^{\nu_j -1} =   -\infty ,}   \\
{ \lim_{x \to \infty } U'_{i,j} (x) + \frac{\kappa_j \nu_i \eta_i(w_j+ \mu_j)}{w_i+ \mu_i} \left(x \right)^{\nu_j -1} = \infty.}\\
\lim_{x \to {f^{min}_i}^+ } U'_{i,j} (x) = -\infty, \lim_{x \to \infty } U'_{i,j} (x) = 0.
\end{align}
Next, according to (\ref{Udiv2}), there is only one solution in the interval $x \in \left[f^{min}_i, + \infty \right)$ for the equation $\frac{\partial \mathcal{L}}{\partial x_{i,j}} =0$, $ i \neq j$, which can be formulated as the following transcendental equations:
\begin{align}
U'_{i,j}(x) +  {\frac{\kappa_j \nu_i \eta_i(w_j+ \mu_j)}{w_i + \mu_i} \left(x \right)^{\nu_j -1} }&= {-\frac{\eta_iv_j}{w_i+ \mu_i}}, j \neq i, j \in \mathcal{N} \label{eqn}, \\
U'_{i,0}(x) &={ -\frac{\eta_iv_0}{w_i+ \mu_i}}.\label{eqn1}
\end{align}
For simplicity, denote the solution to (\ref{eqn}) in the interval $x \in \left[f^{min}_i, + \infty \right)$ as $\Gamma_{i,j}$, and denote the optimal solution to Problem (\ref{pro3}) as $(a_{i,j}^*, x^*_{i,j})$.
Obviously, if $x_{i,j}^*=0$, then $a_{i,j}^* =0$, which is due to the constraints that 
$x_{i,j} \in [a_{i,j}f_{i,j}^{D}, a_{i,j}f_{i,j}^{U}]$.
In the following analysis, we first consider the case that $x_{i,j}^* =0$, and then consider the case that $x_{i,j}^* \neq 0$.

\subsubsection{if $x_{i,j}^* =0$}
First of all, according to Proposition \ref{pro_2}, it is inferred that $ x_{i,k}^* = 0, a_{i,k}^* = 0, \forall  k \in \mathcal{J}_{i}$.
In addition, we define the set $\mathcal{K}_{i}$ as
\vspace{-0.5em}
\begin{equation}\label{Ki}
\mathcal{K}_{i}\!\!= \!\!\left\{j \Big|\mathcal{L}|_{x_{i,j}=0,a_{i,j}=0 } < \!\!\min\!\left\{\!\mathcal{L}|_{x_{i,j}=f_{i,j}^{U},a_{i,j}=1}, \mathcal{L}|_{x_{i,j}=f_{i,j}^{D},a_{i,j}=1}, \mathcal{L}|_{x_{i,j}=\Gamma^*_{i,j},a_{i,j}=1}\right\}\right\}\!\!, j\in \mathcal{M}\!\setminus \!\mathcal{J}_{i}\!\!\!\!
\end{equation}
where $\Gamma_{i,j}^* =  [\Gamma_{i,j}]_{f_{i,j}^{D}} ^ {f_{i,j}^{U}}$. 
Term $y =[x]_a^b$ means that if $ x \geq a$, then $y = a$, if $x \leq b$, then $y = b$, otherwise, $y =x$.
Then, for any device $j, j \in \mathcal{K}_{i}$, we have 
$x_{i,j}^* = 0, a_{i,j}^* = 0, \forall  j \in \mathcal{K}_{i}$.
That is to say, with the current dual variables, task $i$ should \textbf{not} be offloaded to the devices in set $\mathcal{K}_{i}$.

\subsubsection{if $x_{i,j}^* \neq 0$}
Then, we consider the case that $x_{i,i}^* \neq 0$.
In this case, according to (\ref{Lxii}), as the dual variables are non-negative, we have
$\frac{\partial \mathcal{L} }{\partial x_{i,i}}  > 0 , \forall i = j, i \in \mathcal{N}$.
Then, it is inferred that if the task is executed locally, the UE should compute in the least computation speed that can satisfy the delay constraint, i.e., 
${x_{i,i}}^* = a_{i,i}^* f^{D}_{i,i}$.
Furthermore, if $x_{i,j}^* \neq 0$ and $a_{i,j}^* \neq 0$, $\forall i \neq j $, according to the KKT conditions, we can conclude the following conditions:
\begin{equation}\label{div2}
\frac{\partial \mathcal{L}}{\partial a_{i,j}} 
\left\{\begin{array}{llll}
=0, & \text{if } {a_{i,j}}^* \in (0, 1),  \\ 
<0, & \text{if }  {a_{i,j}}^* =1.
\end{array}\right.
\end{equation}
\vspace{-0.75 em}
\begin{equation} 
\left\{\begin{array}{llll}
\frac{\partial \mathcal{L}}{\partial x_{i,j}}(x)\left.|\right._{x \in \left[f_{i,j}^{D}, f_{i,j}^{U}\right]} > 0 
\text{ and }
\mathcal{L}|_{x_{i,j}=0,a_{i,j}=0} \geq \mathcal{L}|_{x_{i,j}=f_{i,j}^{D},a_{i,j}=1}  & \text{if } x_{i,j}^* =f_{i,j}^{D},   \\ 
\frac{\partial \mathcal{L}}{\partial x_{i,j}}(x)\left.|\right._{x \in \left[f_{i,j}^{D}, f_{i,j}^{U}\right]} = 0
\text{ and } 
\mathcal{L}|_{x_{i,j}=0,a_{i,j}=0} \geq \mathcal{L}|_{x_{i,j}=\Gamma_{i,j}^*,a_{i,j}=1}
, & \text{if } {x_{i,j}}^* \in (f_{i,j}^{D}, f_{i,j}^{U}) ,  \\ 
\frac{\partial \mathcal{L}}{\partial x_{i,j}}(x)\left.|\right._{x \in \left[f_{i,j}^{D}, f_{i,j}^{U}\right]} < 0
\text{ and }
\mathcal{L}|_{x_{i,j}=0,a_{i,j}=0} \geq \mathcal{L}|_{x_{i,j}=f_{i,j}^{U},a_{i,j}=1}
, & \text{if }  {x_{i,j}}^* = f_{i,j}^{U}.
\end{array}\right.
\end{equation}

Then we have 
\vspace{-0.75 em}
\begin{equation}
x_{i,j}^* = a_{i,j}^* \Gamma_{i,j}^*, \forall i \neq  j , i\in \mathcal{N},  j  \in \mathcal{M}.  \label{xij}
\end{equation}

To determine the task decision $a_{i,j}$, we define $I_{i,j}$ as 
\begin{equation}
{I_{i,j} = \frac{(w_i+ \mu_i)}{\eta_i}\left( U_{i,j}\left(\Gamma_{i,j}^*\right) - \Gamma_{i,j}^* U_{i,j}'\left(\Gamma_{i,j}^*\right)\right),
 \forall i \neq  j , i \in \mathcal{N}, j \in \tilde{\mathcal{M}}_i,}
\end{equation}
where the set $\tilde{\mathcal{M}}_i$ is defined as
$\tilde{\mathcal{M}}_i =  \mathcal{M}\setminus \{\mathcal{\mathcal{K}}_{i} \cup \mathcal{\mathcal{J}}_{i}\}$.
{In fact, according to Proposition \ref{pro_2} and (\ref{Ki}), $\tilde{\mathcal{M}}_i$ represents the device set that can execute task $i$ successfully.}

Note that $U_{i,j}'\left(x\right)\leq 0$ for all $x \in [f^{min}_i, + \infty]$.
Consequently, according to (\ref{Laij}) and (\ref{Laii}), if $x_{i,i} \neq 0$ and $x_{i,j} \neq 0$, the following inequality always holds
\begin{equation} \label {Lai}
\frac{\partial \mathcal{L} }{\partial a_{i,i}} < \frac{\partial \mathcal{L} }{\partial a_{i,j}} .
\end{equation}

Finally, based on the above analysis, the task decision $a_{i,j}$ is concluded as
\begin{equation} \label{aa}
\left\{\begin{array}{llll}
{a_{i,j}^*}=0, \forall j \in \mathcal{M}, & \text{if } \tilde{\mathcal{M}_i} = \varnothing,\\ 
a_{i,i}^* = 1, {a_{i,j}}^*=0, j \neq i, j \in \mathcal{M}, & \text{if }  i \in \tilde{\mathcal{M}_i},\\ 
{a_{i,k}^*}=1, {a_{i,j}}^*=0, j \neq k, j \in \mathcal{M}, & \text{else}, k = \arg \underset{ j \in \tilde{\mathcal{M}}_i }{\min} I_{i,j}.
\end{array}\right.
\end{equation}

{ 
To determine the offloading decision, we have three possible cases: 
First, if the feasible device set $\tilde{\mathcal{M}}_i$ is empty, this task cannot be accomplished successfully anywhere so task $i$ should not be assigned, i.e., $a_{i,j} =0, \forall j \in \mathcal{M}$.
Second, when $\tilde{\mathcal{M}}_i \neq \varnothing$, UE can process its own task in time, i.e., $i \in \tilde{\mathcal{M}_i}$.
According to (\ref{Lai}), this task should be compute locally, i.e., $a_{i,i} =1$.
Otherwise,  as the task can be offloaded to only one device (constraint $C2$), if $I_{i,j}$ for different $j\in \tilde{\mathcal{M}_i}$ are all different, then only the device with the smallest $I_{i,j}$ should be assigned with task $i$.}

The transmit power can be readily obtained according to (\ref{PT}).
Note that the value of the dual variables $\mu_i$ and $v_ j$ can be determined by the sub-gradient method.
The updating of $\mu_i$, and $v_ j$ in the $(t +1)$-th iteration are
\begin{align}
\mu_i^{(t+1)}=& \left[\mu_i^{(t)}  + \theta_i^{(t)}{ \left(\sum_{j \neq i, j\in \mathcal{M}}   \frac{a_{i,j}^{(t)} w_{i,j} }{\eta_i} U_{i,j}\left(\Gamma_{i,j}^{*(t)}\right) + \kappa_i \sum_{k \in \mathcal{N}} a_{k,i}^{(t)} \left(\Gamma_{k,i}^{*(t)}\right)^{\nu_i} - p^{m}_i \right)}\right]^{+} \!\!\!, i \in \mathcal{N}, \label{Var1}   \\
v_ j^{(t+1)}=& \left[v_ j^{(t)}+ \zeta_j^{(t)} \left(  \sum_{i \in \mathcal{N}}x_{i,j}^{(t)} - f_{j}^{max} \right)\right]^{+}, j \in \mathcal{M}, \label{Var2}
\end{align}
where $[a]^+ = \max\{0,a\}$, and $\theta_i^{(t)}$ and $\zeta_j^{(t)}$ are the positive step-sizes in $t$-th iteration. 
According to \cite[Proposition 6.3.1]{bertsekas1999nonlinear}, the sub-gradient method converges to the optimal solution to Problem (\ref{pro3}) for sufficiently small  step-sizes.
Overall, the above analysis is summarized in Algorithm 1.

\begin{algorithm}
	\caption{Integer Constraints Relaxation Based Iterative (ICRBI) Algorithm}
	\begin{algorithmic}\label{alg1}
		\STATE Initialize ${x_{i,j}}^{(0)}=0,{a_{i,j}}^{(0)}=0, \forall i \in \mathcal{N}, j \in \mathcal{M}$ and the precision parameters $\epsilon$;\\ 
		\STATE Initialize $\mu_i^{(0)}$, $v_ j^{(0)}$, $ \theta_i^{(0)} $, and $\zeta_j^{(0)} $, $\forall i \in \mathcal{N}, j \in \mathcal{M}$; \\
		\REPEAT
		\FOR{$i \in \mathcal{N}$,$j \in \mathcal{M}$}
		\STATE  Calculate ${x_{i,j}}^{(t)}$ and ${a_{i,j}}^{(t)}$ according to (\ref{xij}) and (\ref{aa}), respectively;
		\ENDFOR
		\STATE Update $\mu_i^{(t)}$, $v_ j^{(t)}$ according to (\ref{Var1}) and (\ref{Var2}), respectively;
		\STATE Update the value of the objective function according to (\ref{obj3}) ;
		\UNTIL  $|\mathcal{C}^{(t)} - \mathcal{C}^{(t-1)}|< \epsilon$
	\end{algorithmic}
\end{algorithm}

{The proposed ICRBI algorithm is based on the Lagrangian dual subgradient method, and its computational complexity consists of two parts: the optimization of the Lagrangian dual with given dual variables and finding the dual variables by the sub-gradient method.
For the first part, the main complexity lies in solving the transactional equations (\ref{eqn})-(\ref{eqn1}), as the rest parts are provided with closed-form expressions.
Suppose that the Newton method is adopted and the required number of iterations in the worst case is denoted by $T_{N}$.
The total number of transactional equations is $N(N+1)$.
For the second part, the complexity is caused by finding the optimal dual variables. 
In Problem (\ref{pro3}), the number of dual variables that need to be updated is $2N+1$.
Let $T_{D}$ be the number of subgradient updates needed.
Then, we can conclude that the total complexity of the ICRBI algorithm is 
$\mathcal{O}\left(N(N+1)T_{N}(2N+1)T_{D}\right) =\mathcal{O}\left(N^3T_{N}T_{D}\right)$.}

\vspace{-1 em}
\section{Heuristic Task Matching}

Although the ICRBI algorithm can give efficiently solve Problem (\ref{pro1}), its computation complexity can still be relatively high.
In this section, we propose a heuristic low-complexity matching approach to obtain a feasible solution while still preserving reasonable performance.

\vspace{-1em}
\subsection{Many-to-one Matching}

To determine the task offloading, we model the offloading decision problem as a many-to-one matching between the task set $\mathcal{N}$ and the set of computing devices $\mathcal{M}$.
Let $\Omega$ denote the matching function, i.e., $\Omega(k) = I$ means that task $k\in\mathcal{N}$ is offloaded to device $I\in \mathcal{M}$ for execution under the current matching $\Omega$.  

For the offloading decision, each task aims to identify the device with the least power cost for offloading and computing.
Then, we define the power cost function for a task $k \in \mathcal{N}$ computed by the device $I \in \mathcal{M}$ under the matching $\Omega$ as 
\begin{equation} \label{psi}
\Psi_{k,I}^{\Omega}(f_{k,I})={
\left\{\begin{array}{llll}
\frac{w_k}{\eta_k}  U_{k, I}(f_{k,I})+w_I\kappa_{I}(f_{k,I})^{\nu_I} - \phi_k , &  I \in \mathcal{M}\setminus \{0,k\},  \\ 
\frac{w_k}{\eta_k}  U_{k, I}(f_{k,I}) - \phi_k , & I = 0,\\
w_I\kappa_{I}(f_{k,I})^{\nu_I} - \phi_k , & I = k.
\end{array}\right.}
\end{equation}

Our target is to find a matching so that the system cost can be minimized. 
Then, a matching game is defined by the sets $(\mathcal{N},\mathcal{M})$ and their preference relations, which allows each task $k \in \mathcal{N}$ to rank their preferred devices for execution in $\mathcal{M}$. 
For a task $k \in \mathcal{N}$ with two matchings $\Omega(k) = I$ and $\tilde{\Omega}(k) = J$, $I, J \in \mathcal{M}, I \neq J$, its preference relation $\succ_k$ is defined over the set of devices $\mathcal{M}$ such that 
\begin{equation} \label{prek}
(I, \Omega)\succ_k (J, \tilde{\Omega}) \Leftrightarrow  
\Psi_{k,I}^{\Omega}(f_{k,I}) < \Psi_{k,J}^{\tilde{\Omega}}(f_{k,J}).
\end{equation} 
According to (\ref{psi}) and (\ref{prek}), the power cost $\Psi_{k,I}^{\Omega}(f_{k,J})$ depends on the computation speed $f_{k,I}$ provided by this device. 
However, the computation speed $f_{k,I}$ is coupled with that of the other tasks which are matched to the same device $I$.
Consequently, the preferences of tasks on the devices depend on the current matching that already exists. 

\vspace{-1em}
\subsection{Task Matching Preference List}
\begin{definition} \label{def_5}
Under the current matching $\Omega$, the preference list of task $k$ denoted by $\mathbb{P}_{k}^{\Omega}$ is a list of devices $I, I \in \{\mathcal{M} \setminus\mathcal{\tilde{J}}^{\Omega}_{k}\}$, which are ordered from the most favourite one to the least preferred one according to its preference relation $\succ_k$.
That is, 
\begin{equation}
\Psi_{k,\mathbb{P}_{k}^{\Omega}(1)}^{\Omega}(f_{k,\mathbb{P}_{k}^{\Omega}(1)}) < 
\cdots<
\Psi_{k,\mathbb{P}_{k}^{\Omega}(C_k^{\Omega})}^{\Omega}(f_{k,\mathbb{P}_{k}^{\Omega}(C_k^{\Omega})}),
\end{equation}
where $C_k^{\Omega}$ denotes the cardinality of $\mathbb{P}_{k}^{\Omega}$, and $\mathcal{\tilde{J}}^{\Omega}_{k}$ denote the set of devices that cannot successfully execute task $k$ under matching $\Omega$.
\end{definition}
\vspace{-1em}

To obtain $\mathbb{P}_{k}^{\Omega}$ for task $k$, we first identify the available computation resources that each device can provide to task $k$. 
With matching $\Omega$ in place, the set of tasks that are already matched to device $I$ is denoted by $\mathcal{X}_{I}$, i.e., $\Omega(k') = I, \forall k' \in \mathcal{X}_{I}$.
For the matched task $k', k' \in \mathcal{X}_{I}$, the computing frequencies provided by device $I$ for task $k'$ is denoted as $f_{k',I}$.
Then, the remaining computing capability in device $I, I \in \mathcal{M}$ is 
\begin{equation} \label{FLeft}
\tilde{f}_I^{\Omega}=f_{I}^{max}-\sum_{k'\in \mathcal{X}_{I}}f_{k',I}.
\end{equation}
In addition, the remaining power budget in device $I, I \in \mathcal{N}$ is 
\begin{equation} \label{Pleft}
\tilde{p}_I^{\Omega} =  \left\{\begin{array}{llll}
p^{m}_I - \kappa_i \sum_{k'\in \mathcal{X}_{I}}(f_{k',I})^{\nu_I}  & \text{if } {\Omega(I)}= I, \\
p^{m}_I - \kappa_i \sum_{k'\in \mathcal{X}_{I}}(f_{k',I})^{\nu_I} - \frac{1}{\eta_I}\tilde{U}_{I,\Omega(I)}\left(f_{I,\Omega(I)}\right),& \text{if } {\Omega(I)} \neq I,f_{I,\Omega(I)} \neq 0 .
\end{array}\right.
\end{equation}
where $\Omega(I)$ denotes the device that task $I$ is offloaded to. 

Based on (\ref{FLeft})-(\ref{Pleft}) and the proof of Proposition \ref{pro_2},  to successfully execute task $k$, with the matching $\Omega$ in place, the upper bound $f_{k,I}^{U}$ and the lower bound $f_{k,I}^{D}$ of the computation speed provided by device $I$ are modified to
\begin{align}
\!\!\!\!\tilde{f}_{k,I}^{U} = 
\left\{\begin{array}{llll}
\!\!\!\min \left\{\tilde{f}_I^{\Omega}, \left(\frac{\tilde{p}_I^{\Omega}}{\kappa_I}\right)^{\frac{1}{\nu_I}}\right\}, & \!\!\!\!\forall I \in \mathcal{N},\\ 
\!\!\!\tilde{f}_0^{\Omega}, &\!\!\!\! I =0,
\end{array}\right.,
\tilde{f}_{k,I}^{D} = 
\left\{\begin{array}{llll}
\!\!\!\frac{F_k}{T_k - \frac{D_k}{\tilde{R}_{k,I}^{max}}},&\!\!\!\!\forall I\neq k, I \in \mathcal{M}, \\
\!\!\! f^{min}_k ,&\!\!\!\!I= k,
\end{array}\right.
\label{FupFdn} 
\end{align}
where $\tilde{R}_{k,I}^{max} = \log_2\left(1+ \frac{h_{k,I}\eta_k\tilde{p}_k^{\Omega}}{\sigma^2}\right)$ is the maximum transmit rate of device $k$ with its remaining power budget $\tilde{p}_k^{\Omega}$.
According to Proposition \ref{pro_2}, based on the updated $\tilde{f}_{k,I}^{U}$, $\tilde{f}_{k,I}^{D}$ in (\ref{FupFdn}) and $\tilde{R}_{i,j}^{max}$, the infeasible device set $\tilde{\mathcal{J}}_{k}^{\Omega}$ of task $k$ is
$\tilde{\mathcal{J}}_{k}^{\Omega} = \left\{ I \left| \tilde{f}_{k,I}^{D}  \geq \tilde{f}_{k,I}^{U} \text{ or } T_k^{max} \leq \frac{D_k}{\tilde{R}_{k,I}^{max} }, I \in \mathcal{M} \right.\right \}$.

We first consider that task $k$ is offloaded to the MEC server, i.e. $I=0$.
To accomplish more tasks, the computing frequency $f_{k,0}$ is first set to be 
\begin{equation} \label {FminMEC}
f_{k,0}= \tilde{f}_{k,0}^{D}.
\end{equation}
It is worth pointing out that there may be computing resources left in MEC server after every task is matched. 
Therefore, when all the feasible tasks are matched, the remaining computation resources will be further optimized by leveraging the method in Section V.D.

Then, if task $k$ is executed locally, i.e. $k=I$, it is inferred that 
\begin{equation}
f_{k,k} = \tilde{f}_{k,k}^{D},
\end{equation}
as the power consumption for computation increases with the computing frequency.  

Finally, if task $k$ is matched to device $I, I \in \mathcal{M}\setminus\{0, k, \tilde{\mathcal{J}}_{k}^{\Omega}\}$, the computation speed for task $k$ is determined by solving the following problem:
\begin{subequations}\label{pro4}
	\begin{align}
	\underset{f_{k,I}}{\text{min}}\quad
	&  {\frac{w_k}{\eta_k} U_{k,I}\left(f_{k,I}\right)  
	+ w_I\kappa_I(f_{k,I})^{\nu_I} }\label{obj4} \\
	\text{s.t.} \quad
	&  \tilde{f}_{k,I}^{D} \leq f_{k,I} \leq \tilde{f}_{k,I}^{U}.  \label{st4}
	\end{align}
\end{subequations}

According to (\ref{ufdiv}) and (\ref{Udiv2}), it is easy to infer that Problem (\ref{pro4}) is a convex problem.
By taking the first-order derivative of the objective (\ref{obj4}) with respect to $f_{k,I}$, we have
\begin{equation} \label{Ufdiv}
{\frac{w_k}{\eta_k} U'_{k,I} (f_{k,I}) + w_I\kappa_I\nu_I(f_{k,I})^{\nu_I -1} = 0.}
\end{equation}
According to (\ref{Udiv2}), it is inferred that there is only one root to equation (\ref{Ufdiv}), which is denoted by $\gamma$ and can be obtained by the root-finding algorithms.
Then, $f_{k,I}$ is summarized as 
\vspace{-0.5em}
\begin{equation}\label{fkI}
f_{k,I} = 
\left\{\begin{array}{llll}
\tilde{f}_{k,I}^{D}, & \text{if } I \in \{0,k\},  \\ 
{[\gamma]}^{\tilde{f}_{k,I}^{U}}_{\tilde{f}_{k,I}^{D}}, & \text{Otherwise.}
\end{array}\right.
\end{equation}
With the computing frequencies given in (\ref{fkI}), the corresponding cost function $\Psi_{k,I}(f_{k,I}^*,\Omega)$ and the preference lists defined in Definition 1 are readily obtained.

\vspace{-0.5em}
\subsection{Task Matching Ordering Criteria}

{According to (\ref{aa}), the local computation should always be given the priority.
Let $\mathcal{K}_L$ denote the set of tasks that can be executed locally, and it can be expressed as
\begin{equation}\label{KL}
\mathcal{K}_L= \left \{ i \Big| \frac{F_i}{T^{max}_i} \leq f_i^{max}\right\}.
\end{equation}
Then, we propose two different criteria to determine the matching order for task $k \notin \mathcal{K}_L$,} of which the priorities are to maximize the number of accomplished tasks and to minimize the power costs, respectively. 
In the following, for simplicity, with a current matching $\Omega$, the set of tasks that are feasible ($\mathbb{P}_{k}^{\Omega} \neq \varnothing$) but unmatched is denoted as $\mathcal{Y}^{\Omega}$.

\subsubsection{Maximize the Number of Accomplished Tasks}

A ``short'' preference list, i.e., a large $C_k^{\Omega}$, means that the number of feasible devices for executing task $k$ is very limited.
Consequently, since this criterion focuses on improving the number of accomplished tasks, the task with the ``shortest'' preference list should be allocated with the device first.
However, {a situation that can easily occur in practice is that there may exist tasks with the same cardinality of preference list.
In this case, the task first to be matched is selected according to its total cost including the power cost and its penalty.}
Overall, the task to be matched is denoted by 
\begin{equation} \label{RatioS}
k = \arg \min \left\{{ \Psi_{\hat{k}}^{\Omega}- \psi_{\hat{k}}|\hat{k} = \arg \min{C_{k'}^{\Omega}, k' \in \mathcal{Y}^{\Omega}} }\right\}.
\end{equation}

\subsubsection{Minimize Power Cost}

This criterion is focusing on improving power efficiency in the task assignment.
Under this criterion, the task with the minimum power cost should be selected, which is given by
\vspace{-0.75 em}
\begin{equation}\label{PwS}
\hat{k} = \arg \min \left\{  {\Psi}_{k,\mathbb{P}_{k}^{\Omega}(1)}^{\Omega} |k \in \mathcal{Y}^{\Omega}\right\}.
\end{equation} 

It is worth pointing out that the preference lists of tasks should be updated after each task mapping due to the ``externalities'', since the remaining computing resources and power budget in each device change continuously.

\vspace{-1 em}
\subsection{Remaining Computing Resources in the MEC server}

After all the feasible tasks are allocated with devices for execution, there may be remaining computing resources left in the MEC server.
{Note that the remaining computing resources are not enough to support more UEs at this time, as the matching process has ended.} 
As a result, the remaining computing resources can be utilized to decrease the cost of power consumption.

The set of tasks matched to the MEC server is denoted by $\mathcal{X}^{\Omega}_{0}$.
The assigned computation frequency for task $k, k\in \mathcal{X}^{\Omega}_{0}$ is denoted by $f_{k,0}$.
Correspondingly, according to Proposition 1 and (\ref{psi}), the power cost of task $k$ for offloading is denoted by $\Phi_k = \frac{w_k}{\eta_k}  U_{k, 0}(f_{k,0})$.
The remaining computation resources are proportionally distributed to each task, according to  
the user's power consumption.
Then, the revised computation frequency for task $k$ denoted by $f'_{k,0}$ is
\vspace{-0.5em}
\begin{equation}\label{Remain}
f'_{k,0} = f_{k,0} + \frac{\Phi_k}{\sum_{k \in \mathcal{X}^{\Omega}_{0}} \Phi_k }\left (f_{0}^{max} - \sum_{k \in \mathcal{X}^{\Omega}_{0}} f_{k,0} \right).
\end{equation}

\begin{algorithm}
	\caption{Heuristic Matching Algorithm}
	\begin{algorithmic}\label{alg3}
		\STATE Initialize the matching $\Omega$ with $\Omega(k) = \varnothing$, $\forall k \in \mathcal{N}$; 
		\STATE Determine the set of tasks that should be computed locally $\mathcal{K}_L$;
		\STATE Initialize the unmatched feasible task set $\mathcal{Y}^{\Omega} = \mathcal{N}-\mathcal{K}_L$;
		\STATE Each task $k \in \mathcal{Y}^{\Omega}$ initializes its preference list;
					
		\REPEAT
		\STATE  Determine task $k$ to be matched according to (\ref{RatioS}) and (\ref{PwS});
		\STATE  Match task $k$ to its most preferred device, and remove task $k$ from the unmatched set $\mathcal{Y}^{\Omega}$;
		\STATE  Each task in set $\mathcal{Y}$ updates its preference list, and update the task set $\mathcal{Y}^{\Omega}$;
		\UNTIL  set $\mathcal{Y}^{\Omega}$ is empty;
		\STATE  Revise the computation frequency allocation of the MEC server according to (\ref{Remain})
	\end{algorithmic}
\end{algorithm}

\textbf{Remark}:
According to Algorithm \ref{alg3}, a feasible task is assigned with a device for task computation in each iteration.
{Assume that task $i$ is offloaded to device $j$ in the $(t)$-th iteration.
Then, the transmit power for task $i$ is deducted from the power budget of device $i$, and the computation speed $f_{i,j}$ is subtracted from the total computation capability of device $j$. 
As the number of iterations increases and offloading decisions are being continuously made, the available power and computing resources in each device also continuously decrease.}

{For the proposed heuristic matching algorithm, the complexity lies in the preference list for each task, where the root-finding algorithm is applied to solve Equation (\ref{Ufdiv}).
	Let $T_{M}$ denote the number of iterations required for the root-finding algorithm.
	In the worst case, the number of equations for each user to solve is $N$.
	In addition, in the worst case, the number of iterations needed for the matching is $N-1$.
	Then, we can conclude that the complexity of heuristic matching algorithm is 
	$\mathcal{O}\left(N(N-1)T_{M}\right) =\mathcal{O}\left(N^2T_{M}\right)$.}

\section{Decentralized Algorithm and Complexity Analysis}

{
In the ICRBI algorithm, the decision is made by collecting all UE's computation requirements and their channel state information at the MEC server.
In the heuristic matching algorithm, the matching order of tasks is determined based on the centralized ordering of the tasks' characteristics.
To avoid the centralized  coordination required in the proposed ICRBI/heuristic algorithms and further reduce the complexity, we consider the following decentralized algorithm.}

\vspace{-1em}
\subsection{Decentralized Algorithm}
{The proposed distributed algorithm consists of three steps: 1) Determine the tasks for local computing; 2) Determine the tasks for MEC offloading; 3) Gale-Shapley matching to execute the unmatched tasks. The detailed procedure is presented as follows. }

{Step 1: The set of tasks that are executed locally denoted by $\mathcal{K}_L$ is determined by (\ref{KL}).}

{Step 2: For task $k \notin \mathcal{K}_L$, the required computation frequency of task $k$ denoted by $f_{k,0}$ is given in (\ref{FminMEC}).}
{In order to complete as many tasks as possible, the MEC server sorts the tasks according to the increasing order of $\{f_{k,0}\}$.
Let $\pi$ denote the resorted order of tasks, for $\pi(k) < \pi(k')$, we have $f_{\pi(k),0} \leq f_{\pi(k'),0}$.
Let $\mathcal{K}_{MEC}$ denote the set of tasks that are executed in the MEC server.
Then, the number of tasks in set $\mathcal{K}_{MEC}$ denoted by $\pi(k)$ is determined by 
$\sum_{i=1}^{\pi(k)} f_{i,0}\leq f_0^{max} < \sum_{i=1}^{\pi(k)+1} f_{i,0}.$
If $\sum_{i=1}^{\pi(k)} f_{i,0} < f_0^{max}$, the remaining computation resources in the MEC server is optimized by leveraging the method given in Section V.D.}

{Step 3: For task $i, i \notin \mathcal{K}_L \cup \mathcal{K}_{MEC}$, we adopt the Gale-Shapley algorithm to obtain their matching \cite{Teo.2001}.
To avoid being rejected, the requested $f_{i,j}$ by task $i$ to UE $j$ is set to the minimum frequency $\tilde{f}_{i,j}^{D}$ given in (\ref{FupFdn}), i.e., $f_{i,j} = \tilde{f}_{i,j}^{D}$.
Then, the preference list $\mathbb{P}_i^{\Omega}$ of task $i$ is generated according to the ascending order of $f_{i,j}$, i.e.,
$f_{i,\mathbb{P}_{i}^{\Omega}(k)} < f_{i,\mathbb{P}_{i}^{\Omega}(k+1)}$.
Different from the heuristic algorithm, in the decentralized algorithm, all the unmatched task send the requested $f_{i,j}$ to their most preferred devices at the same time.}
{Device $j$ sorts the  received requests according to the increasing order of $f_{i,j}$, and the resorted order of tasks in device $j$ is denoted by $\pi_{j}$. 
Then, offers for tasks are determined by the current maximum computation capacity of device $j$, i.e.,} 
\begin{equation}\label{DecFreq}
{\sum_{i=1}^{\pi_{j}(k)} f_{i,j} \leq \tilde{f}_j^{\Omega} < \sum_{i=1}^{\pi_{j}(k)+1} f_{i,j}.}
\end{equation}

{Note that offers will not be taken immediately, as there may exist tasks that require lower computation frequencies.
For the task without any offer, it continues to send requests to its preferred device that has not yet rejected it.
Meanwhile, the devices will sort the newly received requests together with the accepted requests. 
If the new requests ask for lower computation frequencies, the device will reject the old requests, and send offers to the new requests.}
{A task had an offer before may get rejected again.
Then, it continues to send requests to preferred devices that have not rejected it until has the new offer or being rejected by all the devices.}	
{Then, the matching process ends until all tasks have offers or there is no offer from any device.}

\vspace{-1em}
\subsection{Complexity and Overhead Analysis}

{For the decentralized algorithm, the complexity is dominated by the Gale-Shapley matching in Step 3, as Step 1 and Step 2 are expressed in closed-form expressions without iterations.
For the worst case, a UE will be rejected for $N-1$ times in the  Gale-Shapley matching stage, and the number of iterations is $N-1$, but in each step, the calculation is given by the closed-form expression.
Then, the total complexity of the decentralized algorithm is 
$\mathcal{O}\left(N-1\right)$.}

{
The overhead cost for information exchange in the decentralized algorithm occurs in steps 2 and 3.}
{In Step 2, each UE requires its channel state information (CSI) $\{h_{i,0}\}$ to determine $\tilde{f}_{k,0}^{D}$ in (\ref{FminMEC}).
In FDD mode, CSI is estimated by the receiver and then fed back to the transmitter through a feedback link.
The MEC server utilizes the remaining resources according to (\ref{Remain}), which requires task $k, k \in \mathcal{X}_0^{\Omega}$ to send power cost $\Phi_k$, and the computed $f'_{k,0}$ will be sent back from the MEC server to UEs.
Let $N^{M}$ denote the number of tasks in $\mathcal{X}^{\Omega}_{0}$, then the overhead in this step is $2N^{M}+ 2N$.}
{In Step 3, let $N_{u}$ denote the number of unmatched tasks.
To determine $f_{k,I}^{D}$ in (\ref{FupFdn}), task $k$ requires the CSI $\{h_{k,I}\}$ to other UEs.
Then, the overhead of determining the preference list for all unmatched tasks is $(N-1)N_{u}$.
In each iteration, each unmatched task sends its $f_{i,j}$, the received UE will send offers or reject requests, and at most $2N_{u}$ scalars are required for information exchange.
Denote the number of iteration  by $T_{i}^{m}$.
As a result, the total overhead in Step 3 is $2T_{i}^{m}N_{u} + (N-1)N_{u}$.}
{Overall, the overhead cost of the proposed decentralized algorithm is $2T_{i}^{m}N_{u} + (N-1)N_{u} + 2N^{M}+ 2N$, and note that $N_{u} << N$, $N^{M} < N$ and  $T_{i}^{m} \leq N-1$.}

\begin{algorithm}
\caption{{Decentralized Task Matching Algorithm}}
\begin{algorithmic}\label{alg4}
		\STATE 	{Initialize the matching $\Omega$ with $\Omega(k) = \varnothing$, $\forall k \in \mathcal{N}$}; 
		\STATE 	{Determine the set of tasks that should be computed locally $\mathcal{K}_L$;}
		\STATE 	{Determine task set $\mathcal{K}_{MEC}$, and optimize the computation frequencies according to (\ref{Remain}) ;}
		\STATE 	{Initialize the unmatched feasible task set $\mathcal{Y}^{\Omega} = \mathcal{N}-\mathcal{K}_L$-$\mathcal{K}_{MEC}$;}
		\STATE 	{Each task $k \in \mathcal{Y}^{\Omega}$ initializes its preference list};
		
		\REPEAT
		\STATE  {Each task $k$ sends request to its most preferred device that has never rejected it;}
		\STATE  	{Each device compares all the received requests (including the previous offered ones),  sends offers to the tasks that it can support by checking (\ref{DecFreq}) ;}
		\UNTIL  	{set $\mathcal{Y}^{\Omega}$ is empty or no offers/requests can be sent;}
	\end{algorithmic}
\end{algorithm}

{For comparison, the overhead costs for the ICRBI algorithm and the heuristic matching algorithm are also provided.
In the ICRBI algorithm, the CSI between all UEs needs to be collected in the central controller,  which incurs $N(N-1)$ real scalars corresponding to all the CSI. 
In addition, as the penalty $\{\phi_i\}$ is set by the central controller, solving Problem (\ref{pro3}) needs the requirements of tasks $\{F_i, D_i, T_{i}^{max}\}$, coefficients for power costs $\{w_i,\eta_i, \zeta_i\}$, which  incurs the overhead cost of $6N$ real scalars.
Then, computed $\{a_{i,j}, f_{i,j}\}$ need to be sent back to the corresponding transmitters, incurs the further information exchange of $2N$ real scalars.
Hence, the total overhead of the ICRBI algorithm is $8N+ N(N-1)$. }
 
{In the heuristic matching algorithm, to obtain the preference list, the coefficients for power costs $\{w_i,\eta_i, \zeta_i\}$ first should be broadcast to all UEs before iteration, which incurs the overhead of $3N$ real scalars.
Then, in each iteration, UEs and the MEC server need to broadcast their current $\{\tilde{p}_I^{\Omega},\tilde{f}_I^{\Omega}\}$ to others, which incurs the overhead of $2N+1$ real scalars.
Then, to determined $\title{f}_{k,I}^U$, $\title{f}_{k,I}^D$ and solve Problem (\ref{pro4}) ,
each unmatched task needs the CSI $\{h_{k,I}\}$, which costs at most $N$ real scalars for each task. 
Let $N_{h}$ denote the number of tasks that cannot be executed locally, so that the number of iterations is $N_{h}$, and the total overhead in heuristic matching iteration is $(N_{h}+ \cdots+2+1 )N = \frac{(N_{h}+1)N_{h}}{2}N$.
In addition, in Section V.D, let $N^{M}$ denotes the number of tasks in $\mathcal{X}^{\Omega}_{0}$, then the overhead for utilizing the remaining resources of the MEC server is $2N^{M}$.
Then, the total overhead cost of heuristic matching algorithm is $2N^{M} + \frac{(N_{h}+1)N_{h}}{2}N + 3N+ (2N+1)N_{h}$.}

\begin{table}
	\vspace{-2em}
	\centering
	\caption{The simulation parameters}
	\label{tab1}
	{\begin{tabular}{|l|l|l|}
			\hline
			Parameters	& Value   \\
			\hline
			Bandwidth $B$	&  2 MHz \\
			\hline
			Noise power density	& $-174$ dBm/Hz \cite{Jiang.2016}  \\
			\hline
			Effective switched capacitance $\kappa_i$ & $10^{-27}$  \cite{You.2017}\\
			\hline
			Computing power constant $\nu_i$ & $3$  \cite{Zhang.2013} \\
			\hline
			Maximum power  $P_{max}$ & $[20-50]$dBm \cite{Jiang.2016,Wu.2017,Zhou.2017} \\
			\hline
			Static Circuit power  $p_i^{cir}$  & $100$mW \cite{Wu.2017,Zhou.2014,Zhou.2017}  \\
			\hline
			Data size of task $D_k$ & $[0.1,0.5]$ Mbits  \\
			\hline
			required computation CPU cycles of tasks $F_k$ & $[1*10^4, 15*10^7]$ cycles \\
			\hline
			Maximum task execution time $T_k$ & $[20, 50]$ ms \\
			\hline
			Maximum CPU capacity of UEs $f_k$ &  [$0.5$ G,$1.5$ G ] cycles/second \\
			\hline
	\end{tabular}}
	\vspace{-2em}
\end{table}

\section{Simulation Results}

In this section, extensive simulation results are presented to show the performance gains achieved by the proposed schemes.
We consider that UEs are uniformly distributed in a $1$ Km $\times$ $1$ Km square cell, where the BS is located in the centre.
{The price $w_i$ is set to 1 for each task unless otherwise specified.
The penalties for tasks $\phi_i$ are evenly distributed in [$\phi_0,\phi_0+10$], where the minimum penalty $\phi_0$ is $40$ unless otherwise specified.}
Most of the simulation parameters are presented in Table \ref{tab1}.
All the results are obtained by averaging over $1000$ random realizations.
For comparison, we consider the computation offloading scheme that the task can either be executed locally or offloaded to the MEC server\cite{8334188}, which is labelled as ``Non-Cope''.
The proposed ICRBI algorithm is labelled as ``ICRBI''.
Furthermore, the proposed heuristic matching approaches with priority to maximize the number of accomplished tasks and to minimize the power costs are labelled as ``MaxTask'' and ``MinPw'', respectively.
{The decentralized algorithm is labelled as the ``DeCentral''.}

\vspace{-1em}
\subsection{The convergence performance}

\begin{figure}
	\begin{minipage}[t]{0.49\textwidth}
		\centering
		\vspace{-2em}
		\includegraphics[width=1\textwidth]{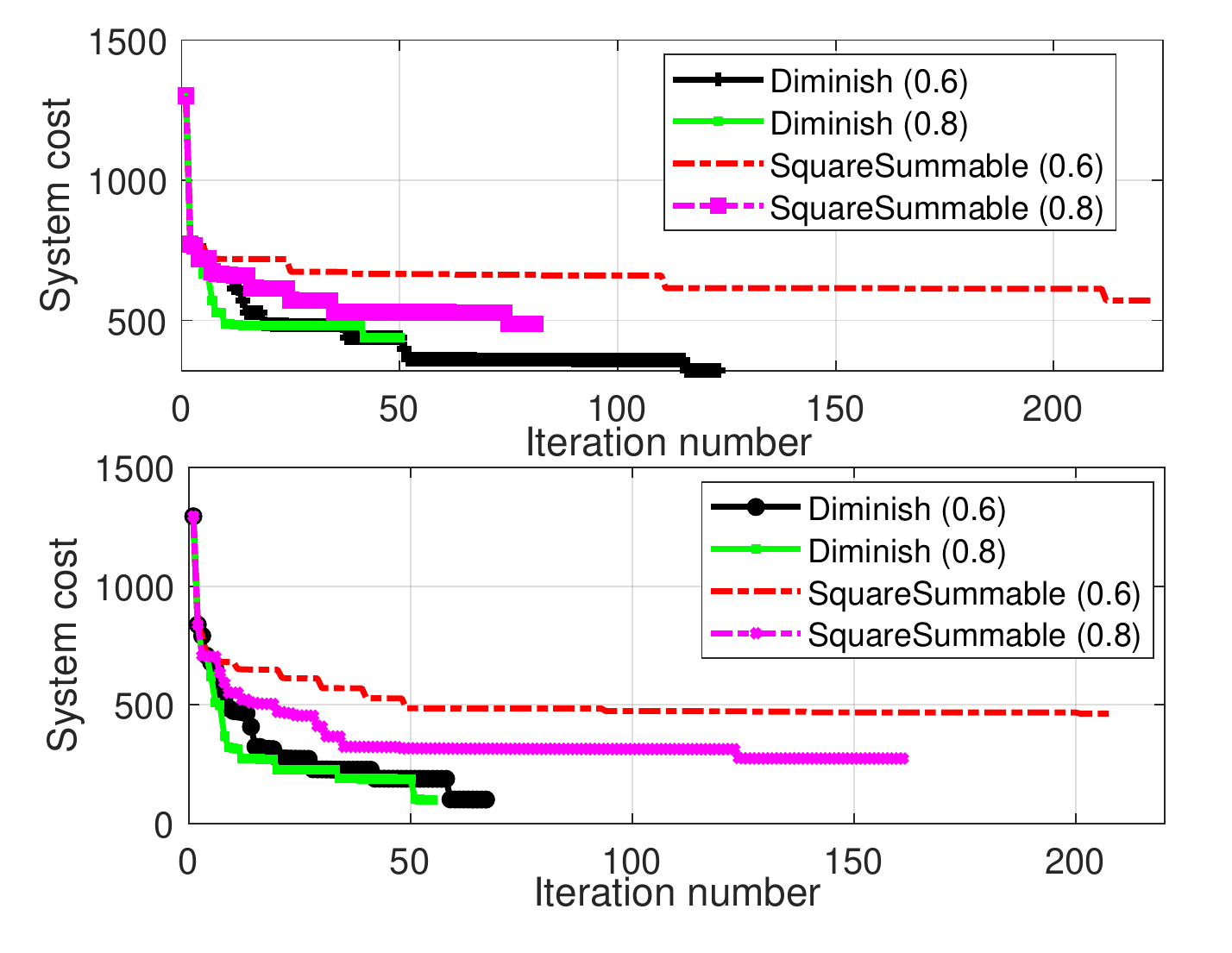}
		\vspace{-3em}
		\caption{{The convergence performance of ICRBI algorithm for two realizations.}}
		\vspace{-2em}
		\label{fig1}
	\end{minipage}
	\begin{minipage}[t]{0.02\textwidth}
	\centering
	\vspace{-3em}
	\includegraphics[width=1\textwidth]{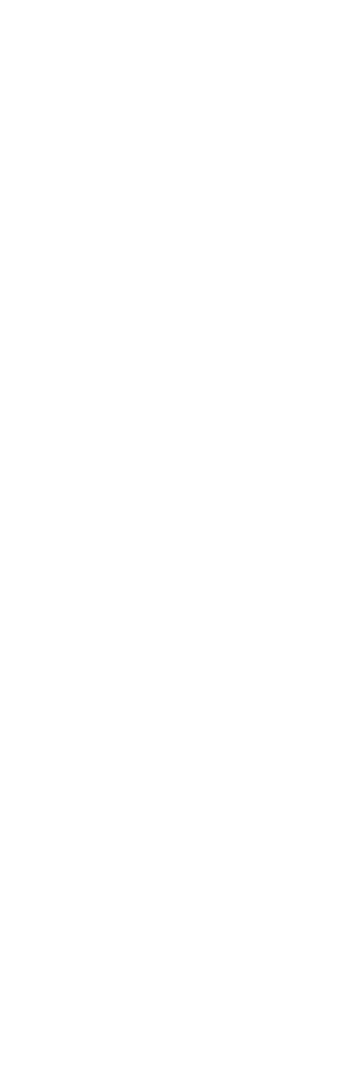}
	\vspace{-1em}
	\end{minipage}
	\begin{minipage}[t]{0.49\textwidth}
		\centering
		\vspace{-2em}
		\includegraphics[width=1\textwidth]{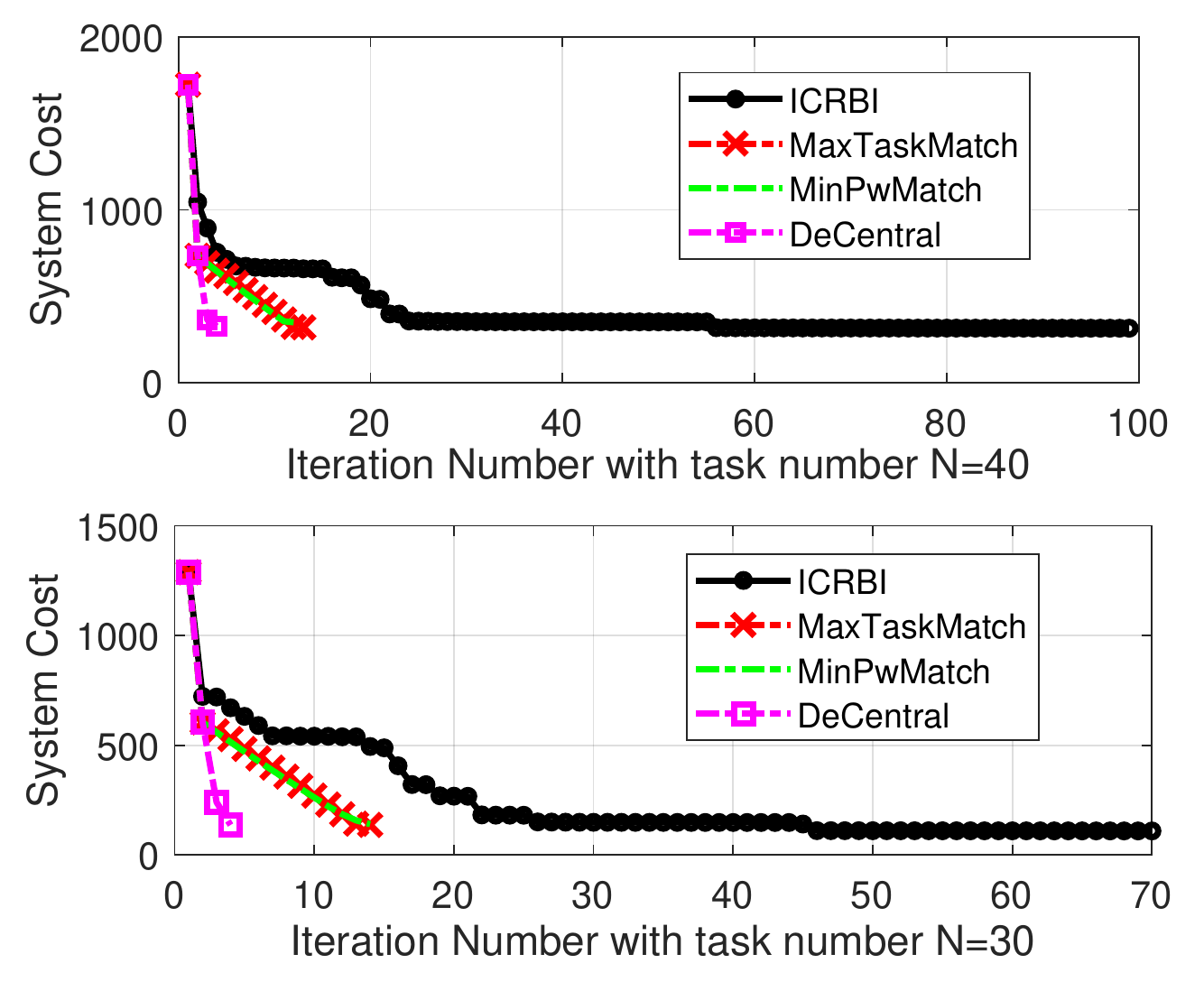}
		\vspace{-3em}
		\caption{{The convergence performance of ICRBI algorithm and the heuristic task offloading approaches.}}
		\vspace{-2em}
		\label{fig2}
	\end{minipage}
\end{figure}

Fig. \ref{fig1} shows the convergence behaviours of the proposed ICRBI algorithm with different step-size updating rules for the dual variable sequences.
{The two sub-figures represent two different system realizations separately, where the users’ locations, requirements of tasks, and fading channels are randomly generated.}
``Diminish$(x)$'' stands for the updating rule $s^l= \frac{x}{\sqrt{l}}$ for the step-size $s$ at the $l$-th iteration; ``Square Summable $(x)$'' stands for the updating rule $s^l= \frac{x}{l}$ for the step-size $s$ at the $l$-th iteration.
In Fig. \ref{fig1}, {$f_{0}^{max}= 5$ G cycles/second, and $N =30$ .}
The system cost monotonically decreases during the initial iterations and then converges for all considered cases in Fig. \ref{fig1}.
In addition, the convergence speed depends heavily on the choice of step-size updating rules.
Compared with ``Square Summable'', the ``Diminish'' updating rule appears to converge faster and be stable across a wider range of the step-sizes for different realizations. 
Therefore, the ``Diminish'' rule is adopted in the following simulations.  

%

Fig. \ref{fig2} illustrates the convergence performance of the proposed algorithms with different numbers of tasks.
The simulation parameters in Fig. \ref{fig2} are the same as those of Fig. \ref{fig1}.
{The decentralized algorithm has the fastest convergence speed but the worst performance in terms of the system cost.
The convergence speed of the heuristic approaches is slightly slower than the decentralized algorithm, but the converged values are better.
Although the convergence speed of ICRBI algorithm is slower, the converged objective value is better than other approaches.}

\vspace{-1em}
\subsection{The impact of system parameters}

\begin{figure}
\vspace{-2em}
	\begin{minipage}[t]{0.5\textwidth}
		\centering
		\includegraphics[width=1\textwidth]{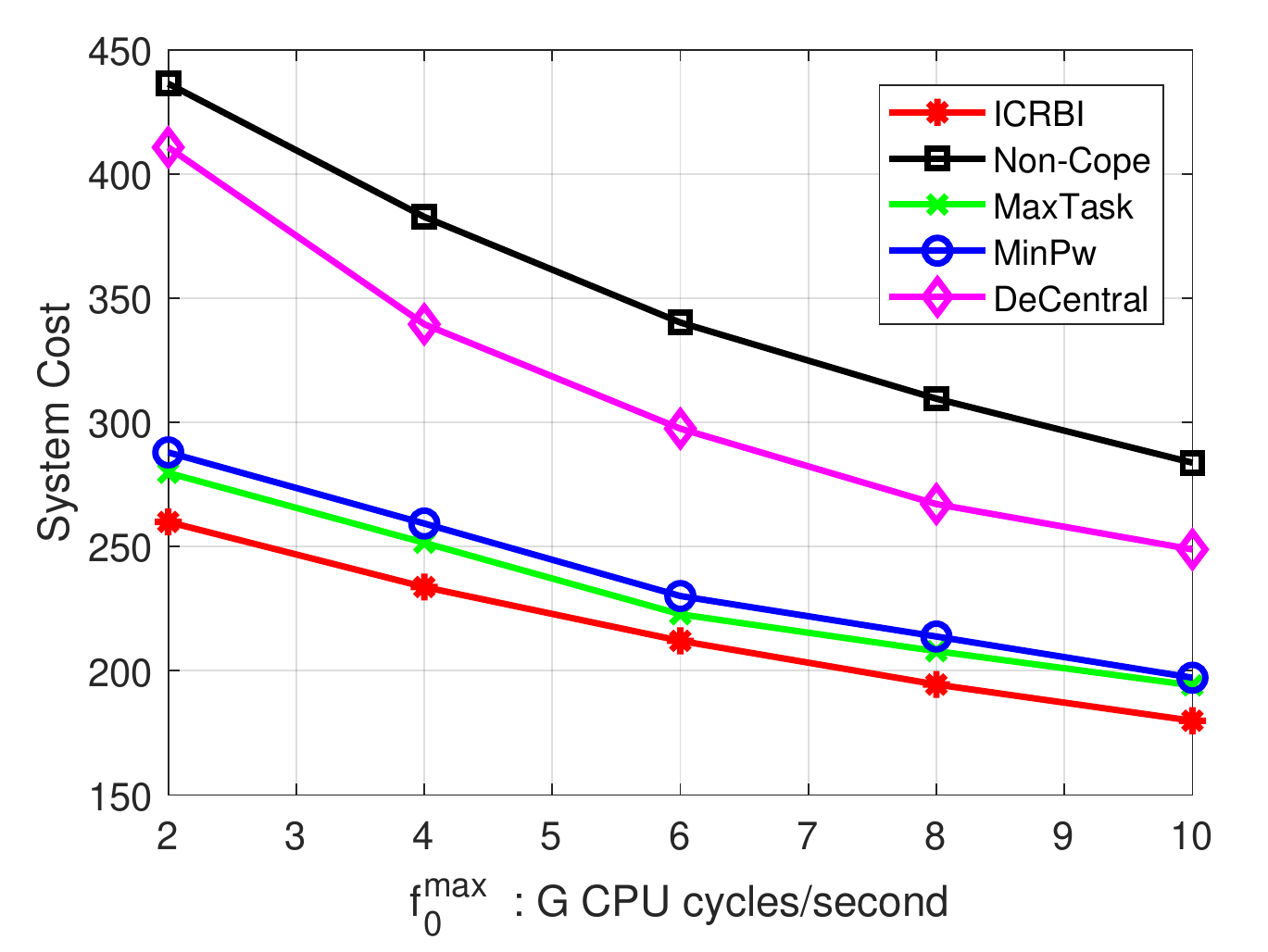}
		\vspace{-3em}
		\caption{{System cost versus the maximum CPU capacity of MEC .}}
		\vspace{-3 em}
		\label{fig3}
	\end{minipage}
	\begin{minipage}[t]{0.01\textwidth}
		\centering
		\includegraphics[width=1\textwidth]{gap.pdf}
		\vspace{-2em}
	\end{minipage}
	\begin{minipage}[t]{0.49\textwidth}
		\centering
		\includegraphics[width=1\textwidth]{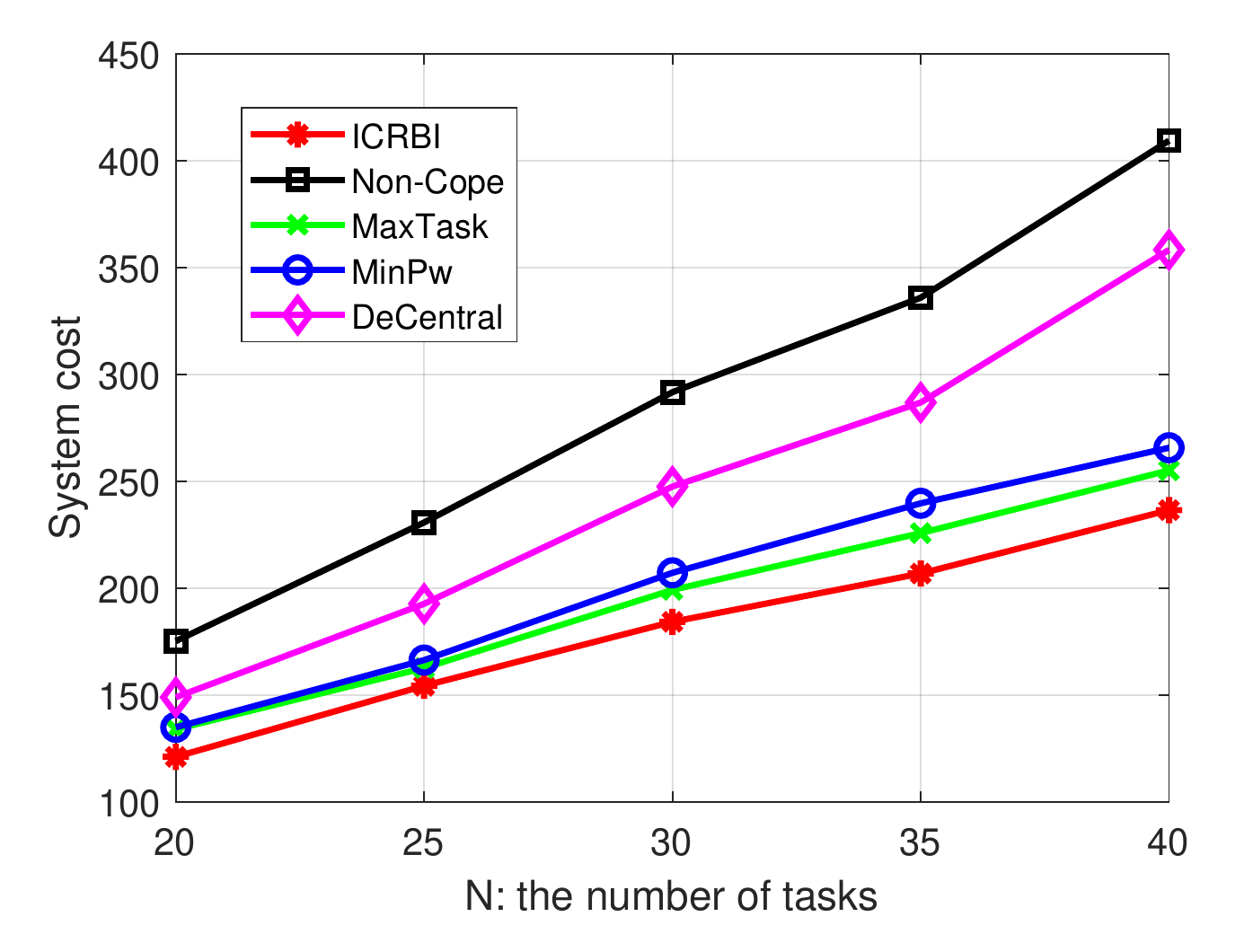}
		\vspace{-3em}
		\caption{{The impact of number of tasks on the total system cost.}}
		\vspace{-3em}
		\label{fig4}
	\end{minipage}
\end{figure}

Fig. \ref{fig3} illustrates the system costs obtained by different algorithms.
In Fig. \ref{fig3}, the number of tasks is $N=30$.
{It is observed that the proposed cooperative computation schemes including the ICRBI algorithm, the heuristic ``MaxTask'' approach, the ``MinPw'' approach and the``DeCentral'' algorithm all outperform the ``Non-Cope'' computation scheme.}
The ICRBI algorithm always achieves the best performance.
The heuristic ``MaxTask'' approach is slightly better than the ``MinPw'' approach due to a large penalty weight (i.e. $\phi_0=40$).
Moreover, system cost $\mathcal{C}$ decreases with the maximum CPU computing capacity of the MEC server.
This implies that more tasks can be accomplished and the power consumption of the mobile devices can be reduced when the computation capability of the MEC server increases.

Fig. \ref{fig4} shows the performance of total system cost versus the number of total tasks with $f_0=8$ G cycles/second.
Clearly, the proposed ICRBI algorithm, ``MaxTask'' approach, the ``MinPw'' approach  and {the ``DeCentral'' algorithm always achieve the better performance than the ``Non-Cope'' scheme.}
In Fig. \ref{fig4}, when the number of tasks increases, the total system cost increases as more power is required for computation.
In addition, the performance gap between the proposed algorithms and the non-cooperative scheme increases with the number of tasks, as more devices can participate in cooperative computing.

\begin{figure}
\vspace{-2em}
	\begin{minipage}[t]{0.49\textwidth}
		\centering
		\includegraphics[width=1\textwidth]{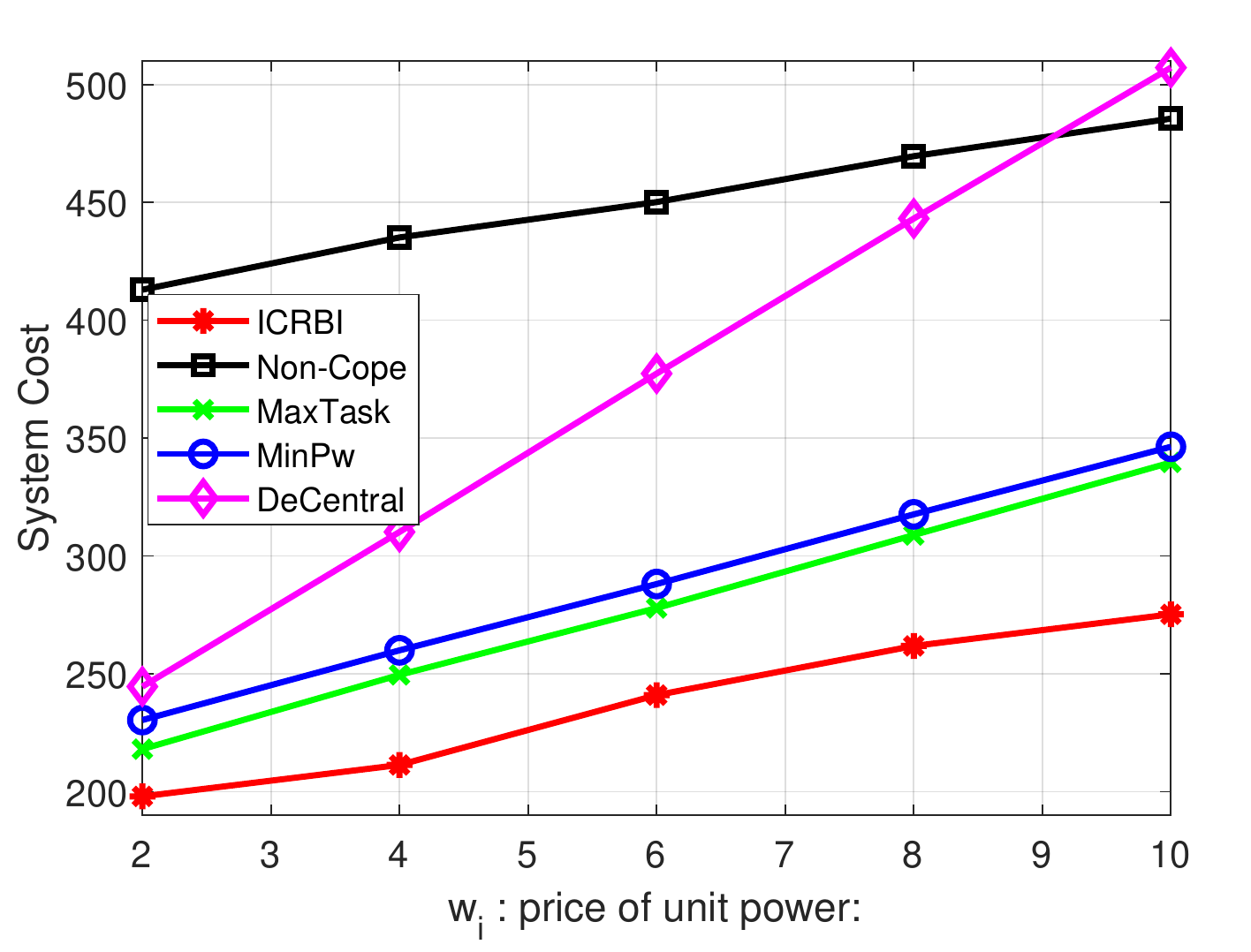}
		\vspace{-3em}
		\caption{{The impact of unit power price on total system cost.}}
		\vspace{-2em}
		\label{fig5}
	\end{minipage}
	\begin{minipage}[t]{0.02\textwidth}
		\centering
		\includegraphics[width=1\textwidth]{gap.pdf}
		\vspace{-3em}
	\end{minipage}
	\begin{minipage}[t]{0.49\textwidth}
		\centering
		\includegraphics[width=1\textwidth]{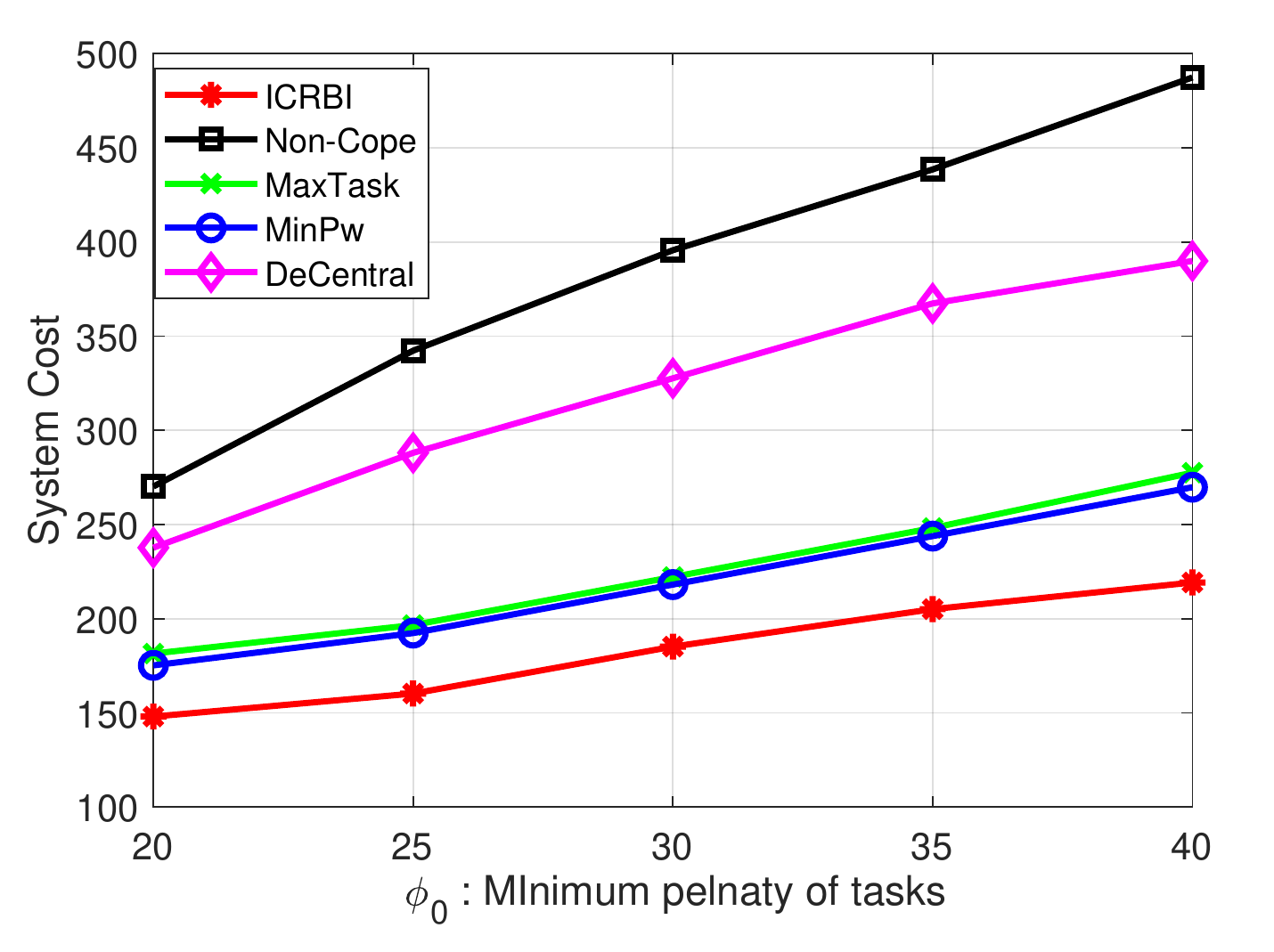}
		\vspace{-3em}
		\caption{{The impact of the penalty on total system cost.}}
		\vspace{-2em}
		\label{fig6}
	\end{minipage}
\end{figure}

Fig. \ref{fig5} shows the impact of unit power price on the total system cost with $f_0=5$ G cycles/second.
As the power price increases, the system cost increases for all considered cases.
This is because the cost of accomplishing one task has increased significantly.
However, the proposed cooperative computation schemes still have lower system costs Compared with the ``Non-Cope'' computation scheme.
{It is interesting to see that the decentralized algorithm is highly affected by the price of unit power, which implies that the central coordination is very important for power saving.}

{Fig. \ref{fig6} illustrates the impact of  task penalty on the system costs.
In Fig. \ref{fig6},  $w_i =5$ and $f_0=5$ G cycles/second.
It is observed that system cost increases with the penalties of tasks.
However, the penalties for accomplishing tasks have a greater influence on system cost obtained by the ``Non-Cope'' scheme and the ``DeCentral''  algorithm than the other proposed algorithms.
The reason is that the ICRBI algorithm, the ``MaxTask'' approach and the ``MinPw'' approach put more efforts on accomplishing tasks when the penalty increases.}

\vspace{-1em}
\subsection{The accomplished tasks}

\begin{figure}
\vspace{-1em}
	\begin{minipage}[t]{0.49\textwidth}
		\centering
		\includegraphics[width=1\textwidth]{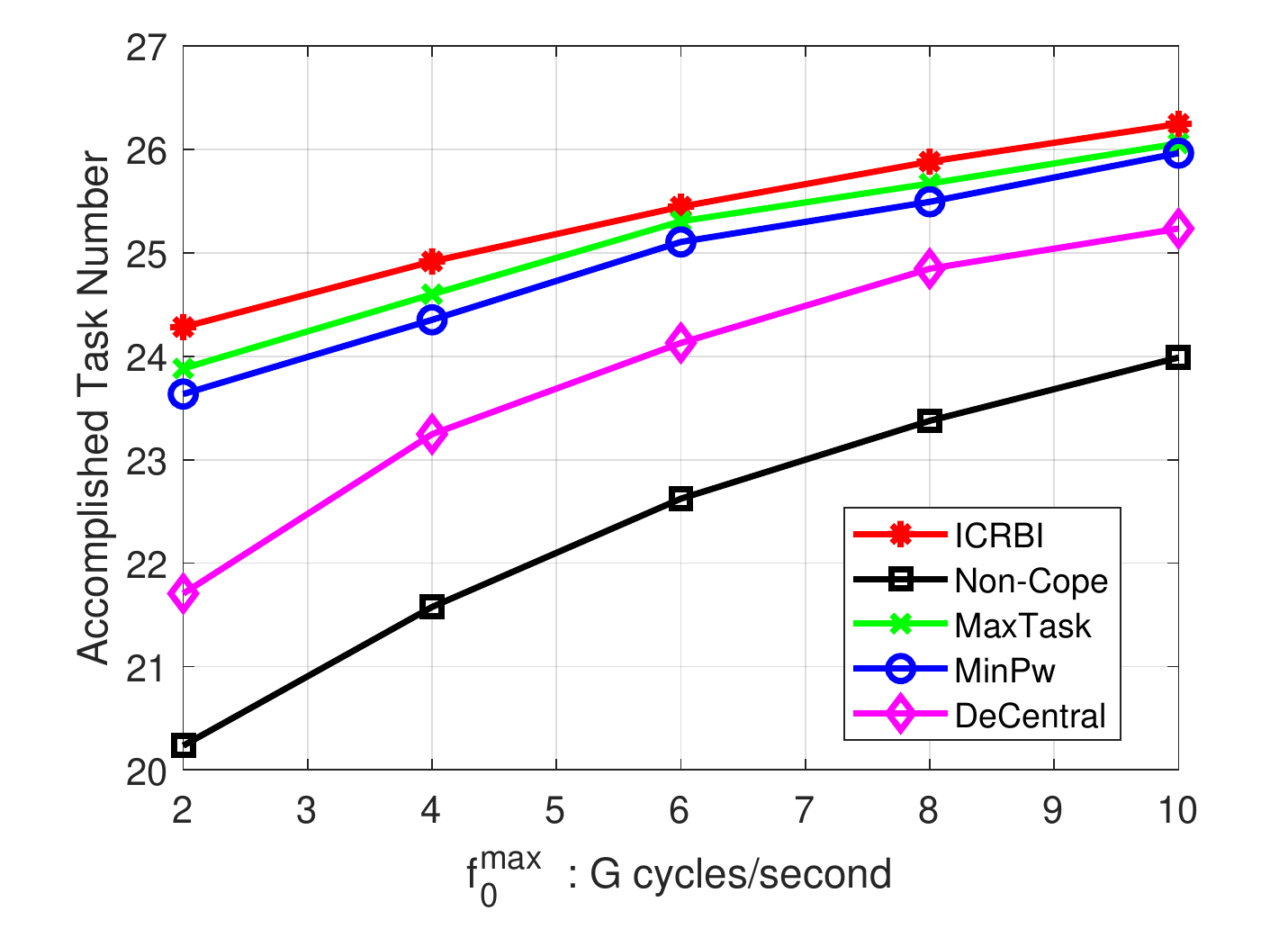}
		\vspace{-3em}
		\caption{{The number of accomplished tasks versus the maximum computation capacity of MEC server.}}
		\vspace{-2em}
		\label{fig7}
	\end{minipage}
	\begin{minipage}[t]{0.02\textwidth}
		\centering
		\includegraphics[width=1\textwidth]{gap.pdf}
		\vspace{-2em}
	\end{minipage}
	\begin{minipage}[t]{0.49\textwidth}
		\centering
		\includegraphics[width=1\textwidth]{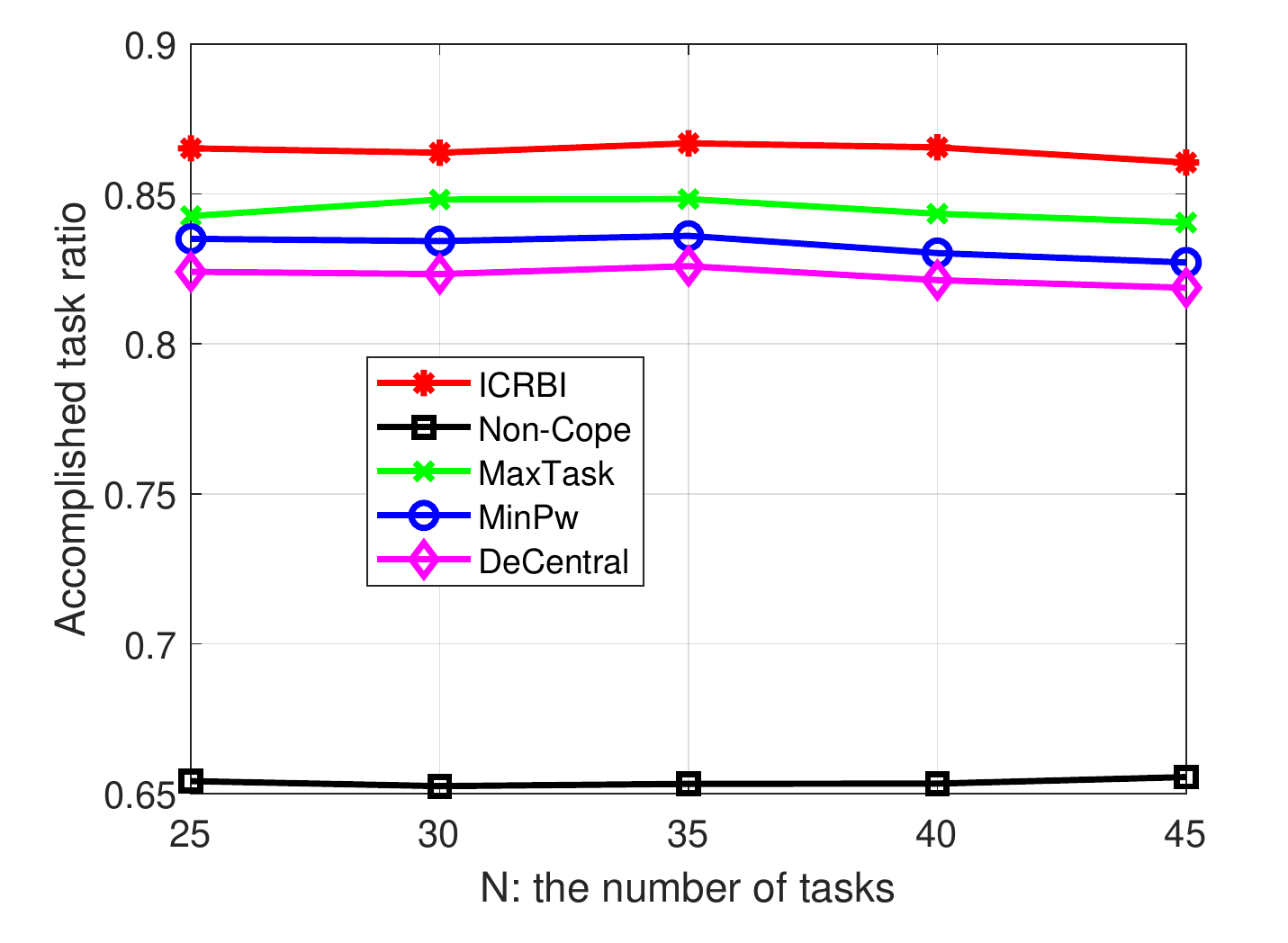}
		\vspace{-3em}
		\caption{{The accomplished task ratios versus the number of tasks.} }
		\vspace{-2em}
		\label{fig8}
	\end{minipage}
\end{figure}

Fig. \ref{fig7} shows the number of accomplished tasks obtained by different algorithms versus the maximum computation capacity of the MEC server with $N =30$.
In Fig. \ref{fig7}, the proposed ICRBI algorithm can achieve the best performance.
The number of accomplished tasks obtained by the ``MaxTask'' approach is larger than those of the ``MinPw'' approach and the ``DeCentral'' algorithm.
This is due to the fact that the matching ordering criterion of ``MaxTask'' approach is to enable more tasks to be accomplished.
{The number of accomplished tasks obtained by the  ``DeCentral'' is lower than the other proposed algorithms, but still higher than the ``Non-Cope'' scheme by exploiting the UE's cooperation.}
When the computing frequency of the MEC server's CPU increases, more tasks can be accomplished, especially for the ``Non-Cope'' scheme.

Fig. \ref{fig8} shows the accomplished task ratios obtained by different algorithms, where the accomplished task ratio is defined as the number of accomplished tasks to the number of total tasks.  
In Fig. \ref{fig8}, the maximum CPU computing capacity of the MEC server is $f_0=5$ G cycles/second.
As expected, the proposed ICRBI algorithm can obtain the largest accomplished task ratio.
The proposed cooperative algorithms still outperform the ``Non-Cope'' scheme by exploiting the resources in the mobile devices.
In addition, when the number of tasks increases, the accomplished task ratio obtained by the proposed algorithms hardly decreases.
This shows that the proposed cooperative computation algorithm can make full use of the computing resources in UEs so that the quality of computation service can be efficiently enhanced.

\vspace{-1em}
\subsection{The power consumption}

\begin{figure}
\vspace{-2em}
	\begin{minipage}[t]{0.46\textwidth}
		\centering
		\includegraphics[width=1\textwidth]{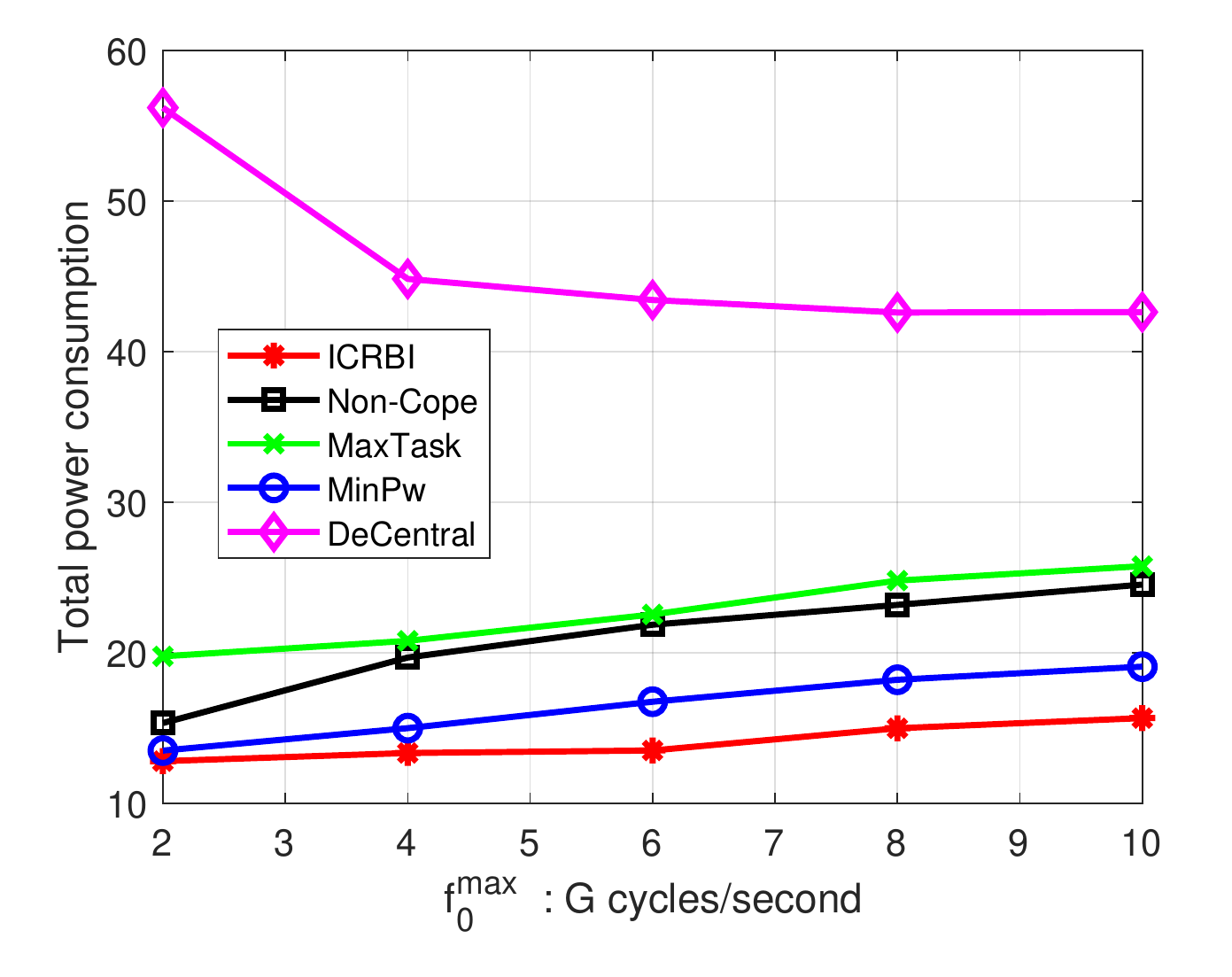}
		\vspace{-3.5 em}
		\caption{{Total power consumption versus the computing capacity of the MEC server's CPU.}}
		\vspace{-2em}
		\label{fig9}
	\end{minipage}
	\begin{minipage}[t]{0.02\textwidth}
		\centering
		\includegraphics[width=1\textwidth]{gap.pdf}
		\vspace{-1em}
	\end{minipage}
	\begin{minipage}[t]{0.46\textwidth}
		\centering
		\includegraphics[width=1\textwidth]{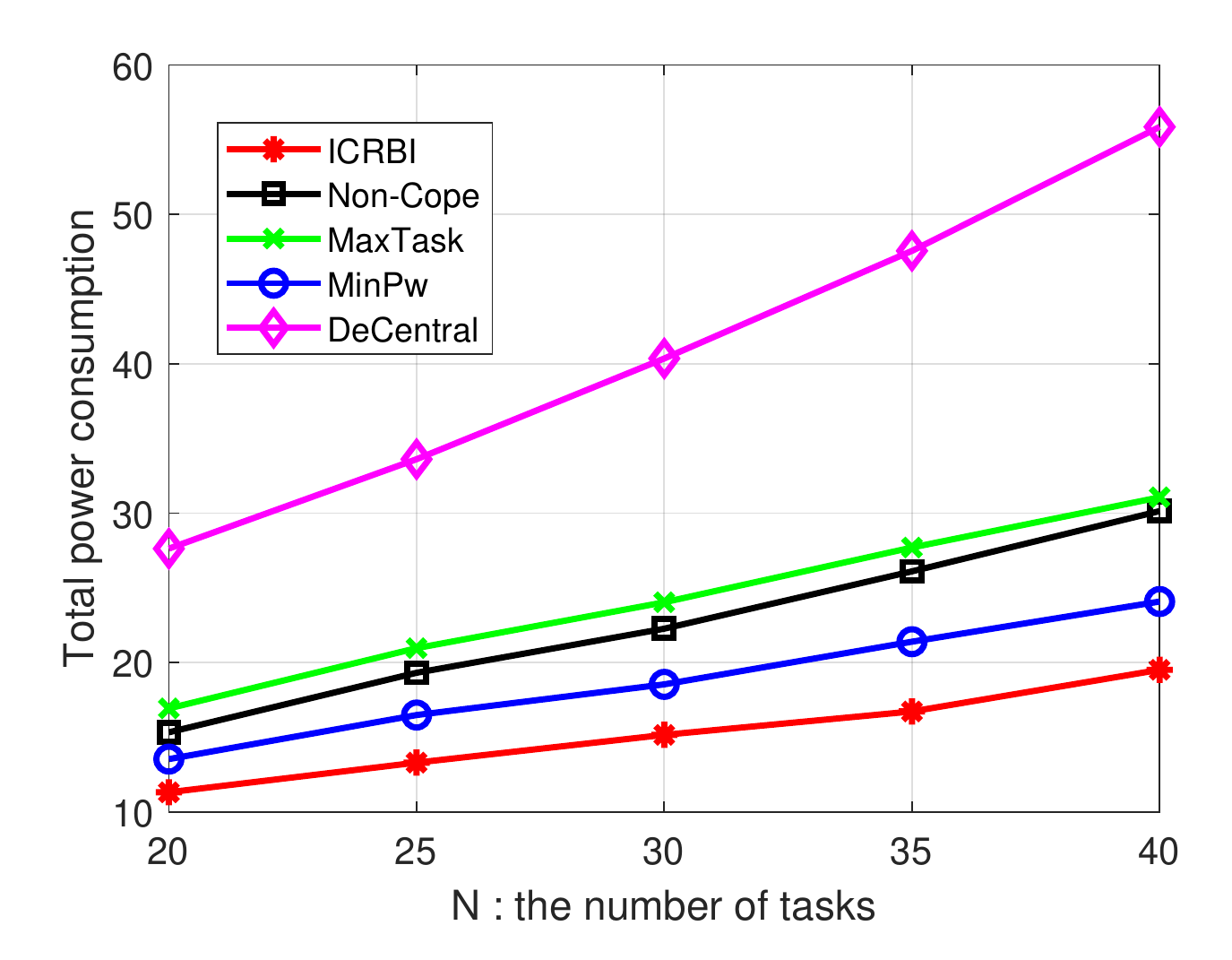}
		\vspace{-3.5 em}
		\caption{{Total power consumption versus the number of tasks.}}
		\vspace{-2.0em}
		\label{fig10}
	\end{minipage}
\end{figure}

Fig. \ref{fig9} shows the total power consumption of UEs versus the computing capacity of the MEC server with $N=30$.
{In Fig. \ref{fig9}, except for the ``DeCentral'' algorithm, the total power consumption of UEs required by the algorithms increases with the CPU capacity as more tasks can be offloaded and accomplished. 
Moreover, the power consumption of the ICRBI algorithm is less than those of all other algorithms, while it can achieve the largest accomplished task ratio.}
In addition, the power consumption of the ``MinPw'' approach is less than that of the ``MaxTask'' approach.
This is because its matching order criterion focuses on maximizing power saving.
It is observed that the power consumption of the ``Non-Cope'' scheme is less than that of the ``MaxTask'' approach due to the small accomplished task ratio.
{Furthermore, the power consumption of the ``DeCentral'' algorithm decreases with the CPU capacity of the MEC server.
This is because more computing tasks can be executed by MEC at this time, so that the power consumption of UEs can be reduced.}

Fig. \ref{fig10} illustrates the total power consumption versus the number of tasks, where the MEC server's computing capacity is $f_0 = 8$G cycles per second.
It is shown that the power consumptions of the cooperative algorithms increase significantly with the number of tasks.
In addition, it is interesting to see that the performance gaps significantly increase with the number of tasks. 
This result comes from the fact that more task computations are executed by the UEs in the cooperative algorithm.
This implies that the cooperative computation algorithms can make full use of D2D communications to reduce the power consumption while still ensuring the large accomplished task ratios.

\section{Conclusions}

{In the considered cooperative computation framework, the offloading decision, the computational frequency, and the offloading power for each UE have been optimized jointly to minimize the system total cost , which consists of the cost charged for UE's power consumption and the penalty caused by the unaccomplished tasks.
Three solutions with different performance and complexity have been provided, i.e., iterative ICRBI algorithm, heuristic matching algorithm and the decentralized algorithm.
According to the theoretical analysis and the simulation results, the following conclusions can be made:}
\begin{itemize}
	\item {By leveraging the proposed algorithms, the cooperative computation framework can sufficiently reduce the system cost compared with the non-cooperative computation scheme, as the computation resources in UEs are exploited. Especially, the ICRBI algorithm can obtain the largest accomplished task ratio meanwhile the power consumption is lowest.} 
	
	\item {The ICRBI algorithm achieves the best performance at the cost of high computation complexity and lower convergence speed. 
	The proposed heuristic matching algorithm can significantly reduce the number of iterations but still achieve acceptable performance.}
	
	\item {Since the ICRBI algorithm and the heuristic matching algorithm are both centralized, a decentralized algorithm is proposed to further reduce the complexity and enable the UE to make decentralized decisions.
	This decentralized algorithm enables the network to provide computing services when it cannot have a central controller, such as emergency communication scenarios in disasters.}
\end{itemize}

\begin{appendices}

	\section{Proof of Proposition \ref{pro_1} }  \label{App1}
	
	{
	We first check that the optimal solution to Problem (\ref{pro1}) denoted by $(\{a^*_{i,j}\},\{f^*_{i,j}\}, \{p^{T*}_{i,j}\})$ is also feasible to Problem (\ref{pro2}).
	As each task can only be offloaded to one device (constraint $C2$),  there are two cases for the offloading decision $\{a^*_{i,j}\}$: 1) $a^*_{i,j} =1$, $i \neq j$; 2) $a^*_{i,j} =1$, $i = j$.}
	
    {For the first case, by observing that the objective function (\ref{obj1}) is an increasing function of $p_{i,j}^T$,  it is inferred that the equality in $C3$ holds for the optimal solution. From  $C3$, we have 
	\begin{equation}\label{gfun}
	r_{i,j} = \frac{D_i f_{i,j}}{T_i^{max}f_{i,j}-F_i} \triangleq G_i(f_{i,j}). 
	\end{equation}
	In addition, according to (\ref{w4}), transmit power $p^{T*}_{i,j}$ can be represented as a function of $r_{i,j}$:
	\begin{equation} \label{PT}
		p^{T*}_{i,j} = \frac{\sigma^2}{h_{i,j}}\left(\exp\left(\frac{\ln(2)}{B}r_{i,j} \right) -1\right) \triangleq H_{i,j}(r_{i,j}).
	\end{equation}
	By combining (\ref{PT}) and (\ref{gfun}), transmit power $p^T_{i,j}$ can be represented as a function of $f^*_{i,j}$:
	\begin{equation}
	 p^{T*}_{i,j} = U_{i,j}(f^*_{i,j}) = \frac{\sigma^2}{h_{i,j}} \left(\exp\left( \frac{\ln 2}{B} \frac{D_i f^*_{i,j}}{T_i^{max}x-F_i}\right)-1\right),
	\end{equation} 
	where $U_{i,j}(x) \triangleq H_{i,j}(G_i(f_{i,j}))$.
	Then, by substituting $p^{T*}_{i,j}$ with $U_{i,j}(f^*_{i,j})$, constraint $C5$ is equivalent to (\ref{Pimax}).
	In addition, as $p^{T*}_{i,j} >  0$, according to the expression of $U_{i,j}(f^*_{i,j}) $, it is verified that constraint (\ref{afmax}) is also satisfied for $\{f^*_{i,j}\}$.
	For the second case, if  $a^*_{i,j} =1$, $i = j$,  then $a^*_{i,j} = 0$, $p^{T*}_{i,j}=0$ and $f^{*}_{i,j}=0$, for all $i \neq j$. }
	
	{In the two cases, the optimal solution $(\{a^*_{i,j}\},\{f^*_{i,j}\}, \{p^{T*}_{i,j}\})$ to Problem (\ref{pro1}) satisfy all the constraints of Problem (\ref{pro2}), so that it is also a feasible solution to Problem (\ref{pro2}). 
	In other words, the optimal objective value of Problem (\ref{pro1}) is no less than that of Problem (\ref{pro2}).}
	
	{Then, we check that the optimal solution to Problem (\ref{pro2}) denoted by $(\{\tilde{a}^*_{i,j}\},\{\tilde{f}^*_{i,j}\})$ is also feasible to Problem (\ref{pro1}).
	There are also two cases : 1) $\tilde{a}^*_{i,j} =1$, $i \neq j$; 2) $\tilde{a}^*_{i,j} =1$, $i = j$.}
	
	{For the first case, we define $\tilde{p}^{T*}_{i,j} =U_{i,j}(\tilde{f}^*_{i,j})$. Substituting $(\{\tilde{a}^*_{i,j}\},\{\tilde{f}^*_{i,j}\}, \{\tilde{p}^{T*}_{i,j}\})$ into the left hand side of constraint $C3$, we have
	\vspace{-0.5em}
	\begin{equation}
	\frac{D_i}{r_{i,j}}+ \frac{F_i}{\tilde{f}^*_{i,j}}= \frac{D_i}{B \log_2\left(1+\frac{\tilde{p}^{T*}_{i,j} h_{i,j}}{\sigma^2}\right)} + \frac{F_i}{\tilde{f}^*_{i,j}}
	= T_i^{max} - \frac{F_i}{\tilde{f}^*_{i,j}} + \frac{F_i}{\tilde{f}^*_{i,j}} = T_i^{max}.
	\end{equation}
	This means that $(\{\tilde{a}^*_{i,j}\},\{\tilde{f}^*_{i,j}\}, \{\tilde{p}^{T*}_{i,j}\})$ satisfies the  constraint $C3$ of Problem (\ref{pro1}).
	 Then, we substitute $\tilde{p}^{T*}_{i,j} =U_{i,j}(\tilde{f}^*_{i,j})$ into (\ref{Pimax}), and the constraint (\ref{Pimax}) is equivalent to the constraint $C5$ of Problem (\ref{pro1}). 
	 For the second case, we have $\tilde{a}^*_{i,j}=0, \tilde{p}^{T*}_{i,j} =0$ for all $i \neq j$. In this case, $(\{\tilde{a}^*_{i,j}\},\{\tilde{f}^*_{i,j}\}, \{\tilde{p}^{T*}_{i,j}\})$ also satisfies all the constraints of Problem (\ref{pro1}).}
	
	{It can be seen that $(\{\tilde{a}^*_{i,j}\},\{\tilde{f}^*_{i,j}\}, \{\tilde{p}^{T*}_{i,j}\})$ are also feasible solutions to Problem (\ref{pro1}), and the optimal objective value of Problem (\ref{pro1}) is no larger than that of Problem (\ref{pro2}).}
	
	{Hence, Problem (\ref{pro2}) is equivalent to Problem (\ref{pro1}).}

\section{Proof of Proposition \ref{pro_2} }\label{App2}

{First, consider that task $i$ is executed locally. 
According to (\ref{Pimax}) and (\ref{ww8}), if UE $i$ consumes all its power to execute task $i$, the obtained computation frequency is
$f_{i,i}^{max} = \left( \frac{p_i^{m}}{\kappa_i }\right)^{\frac{1}{\nu_i}},i \in \mathcal{N}$.
Note that UE $i$'s maximum computation frequency is $f^{max}_i$.
According to the requirements of task $i$, let $f^{min}_i = \frac{F_i}{T_i^{max}}$.
Then, if $f^{min}_i \!> \! f_{i,i}^U = \min \{{f^{max}_i,  f_{i,i}^{max}}\} $, task $i$ cannot be executed locally.}

{Second, we consider that task $i$ is offloaded to the MEC server.
Let $R_{i,0}^{max}$ denote the maximum transmit rate of UE $i$ to the MEC server with its maximum power $p^{m}_i$.
Obviously, if $T_i^{max} < \frac{D_i}{R_{i,0}^{max} }$, task $i$ cannot be offloaded to the MEC server successfully.
Otherwise, to accomplish task $i$, the required minimum computation frequency for the MEC server is
$f_{i,0}^{D} = \frac{F_i}{T_i^{max} - \frac{D_i}{R_{i,0}^{max}}} ,i \in \mathcal{N}$,
where  $R_{i,0}^{max} = B \log_2{\left(1+\frac{h_{i,0}\eta_i }{\sigma^2}p^{m}_i \right)}$.
Then, we can conclude that if $f_{i,0}^{D} > f_{0}^{max} $ or $T_i^{max} < \frac{D_i}{R_{i,0}^{max} }$, task $i$ cannot be executed in MEC server successfully.}

{Finally, we consider the case that $j \neq i, 0, j \in \mathcal{N}$.
As UE $j$ is power-limited by $p_j^{m}$ and computation-limited by $f_j^{max}$ , the maximum provided computation frequency for task $i$  is 
\begin{equation}
f_{i,j}^{U} =  \min \left\{f_{i,j}^{max} = \left( \frac{p_j^{m}}{\kappa_j }\right)^{\frac{1}{\nu_j}} ,f_j^{max} \right\}.\label{Up}
\end{equation}
Moreover, as UE $i$ is also power-limited by $p_i^{m}$, let $f_{i,j}^{D}$ denote the minimum required computation frequency for UE $j$ to accomplish task $i$ and $R_{i,j}^{max}$ represents the maximum transmit rate of UE $i$ to UE $j$. Then, we have 
\vspace{-0.75 em}
\begin{equation}
f_{i,j}^{D}= \frac{F_i}{T_i^{max} - \frac{D_i}{R_{i,j}^{max}}}, \text{ and } R_{i,j}^{max} = B \log_2{\left(1+\frac{h_{i,j}\eta_i }{\sigma^2}p^{m}_i \right)}\label{Dn}.
\end{equation}
Obviously, if $T_i^{max} < \frac{D_i}{R_{i,j}^{max}}$ or $f_{i,j}^{D} > f_{i,j}^{U}$, task $i$ cannot be successfully executed in UE $j$.}

{For notation simplicity, we define 
$f_{i,i}^{D} = f_i^{min},f_{i,0}^{U} = f_0^{max},\text{ and }f_{i,0}^{D} = f_{i,0}^{min}$.
Then, the set of devices that cannot execute task $i$ successfully is summarized as the set $\mathcal{J}_{i}$ given in  (\ref{Jij}).}

\qed

\end{appendices}

\vspace{-1em}

\bibliographystyle{ieeetran}
\bibliography{Refer_0425}

\end{document}